\documentclass[12pt]{article}
\usepackage{graphicx,amssymb,amsmath,amsthm,cite,cases,color}
\usepackage[draft]{hyperref}
\usepackage{mathrsfs}
\usepackage[english]{babel}

\def\bel{\begin{linenomath}}
\def\eel{\end{linenomath}}
\def\be{\begin{equation}}
\def\ee{\end{equation}}
\def\ben{\begin{eqnarray}}
\def\een{\end{eqnarray}}

\def\atr{b^r}
\def\vdel{\boldsymbol{{\delta}}}
\def\vat{\mathbf{b}}
\def\vatr{\mathbf{b}^r}
\def\vhb{\mathbf{h}}

\def\vd{\mathbf{d}}
\def\vs{\mathbf{s}}
\def\jll{\jmath}
\def\jllt{\til{\jll}}

\def\at{b}
\def\atr{b^r}
\def\vAtr{\mathbf{B}^r}

\def\vfr{\mathbf{f}^{\rm{r}}}

\def\CR{\mathrm{CR}}
\def\CRs{\mathrm{CRs}}
\def\SNR{\mathrm{SNR}}
\def\snr{\mathrm{snr}}
\def\SR{\mathrm{SR}}
\def\PRDN{\mathrm{PRDN}}
\def\prdn{\mathrm{prdn}}
\def\dim{\mathrm{dim}}

\def\Dc{\D^{\rm{c}}}
\def\Ds{\D^{\rm{s}}}

\def\CDF{\mathbf{CDF97}}

\def\vbp{\mathbf{b'}}
\def\vbpt{\mathbf{\til{b}'}}
\def\vc{\mathbf{c}}
\def\vc{\mathbf{c}}

\def\vs{\mathbf{s}}
\def\vst{\mathbf{\til{s}}}

\def\dicv{{\cal V}}
\def\dicw{{\cal W}}

\def\vCt{\mathbf{B}}
\def\vC{\mathbf{C}}
\def\vCr{\mathbf{C}^r}

\def\DCT{\widehat{\mathrm{dct}}}

\def\at{b}
\def\atr{b^r}
\def\bpt{\til{b}'}

\def\Proof{\begin{proof}}
\def\EndProof{\end{proof}}

\def\D{\mathcal{D}}
\def\E{\mathcal{E}}

\def\R{\mathbb{R}}
\def\N{\mathbb{N}}
\def\Z{\mathbb{Z}}

\def\vbeta{\boldsymbol{\beta}}

\def\cq{\{q\}}

\def\st{\til{s}}
\def\vst{\mathbf{\til{s}}}

\def\til{\tilde}
\def\wtil{\widetilde}

\def\vr{\mathbf{r}}

\def\vrq{{\vr}{\cq}}
\def\vd{\mathbf{d}}

\def\vc{\mathbf{c}}
\def\vf{\mathbf{f}}

\def\vfa{\vf^{\rm{a}}}

\def\vfq{{\vf}{\cq}}

\def\vft{\til{\mathbf{f}}}

\def\vC{\mathbf{C}}

\def\vW{\mathbf{w}}
\def\vWt{\mathbf{\wtil{w}}}

\def\vd{\mathbf{d}}

\def\Join{\hat{\operatorname{J}}}
\def\Nq{L}

\newcommand{\La}{ \left \langle}
\newcommand{\ra}{\Ra}
\newcommand{\la}{\La}
\newcommand{\Ra}{ \right \rangle}

\newtheorem{proposition}{Proposition}
\newtheorem{corollary}{Corollary}
\newtheorem{remark}{Remark}

\def\rc{\rm{c}}
\def\rs{\rm{s}}

\def\Dc{\D^{\rm{c}}}
\def\Ds{\D^{\rm{s}}}

\newcommand{\Spann}{{\mbox{\rm{span}}}}

\title{Simultaneous Optimized Orthogonal Matching Pursuit with 
	Application to ECG Compression}

\author{Laura Rebollo-Neira,\\
Department of Applied Mathematics and Data Science,\\
Aston University,\\
B3 7ET, Birmingham, UK}

\date{}

\begin{document}
\maketitle
\baselineskip = 1.5\baselineskip
\begin{abstract}
A greedy pursuit strategy which finds a common basis for  approximating a set of similar signals is proposed. The strategy extends the Optimized Orthogonal Matching Pursuit approach to selecting the subspace containing the approximation of all the signals in the set.  The method, called Simultaneous Optimized Orthogonal Matching Pursuit, is stepwise optimal in the sense of minimizing at each iteration the mean square error norm of the signals in the set.  When applied to compression of electrocardiograms, significant gains over other transformation based compression techniques are demonstrated on the MIT-BIH Arrhythmia dataset.

\end{abstract}
\section{Introduction}
Important signals in everyday life such as natural images, 
audio, and electrocardiogram records, 
are in  general highly compressible. This implies that 
the original signal,  available as a large set of numerical 
values, can be transformed into a set of  much  
smaller cardinality or a set containing a large proportion of 
zero values. The transformation, which should not compromise 
the informational content of the data, is frequently
called sparse representation. 
Traditional methods for sparse representation  of signals 
are realized  
by applying an orthogonal transformation and disregarding 
the least relevant points in the transformed domain.  
Subsequently the signal is recovered by means of the
inverse transformation. 
 However, alternative 
transformations, which are not orthogonal but adapted 
to a signal at hand,  
have been shown to render high level of sparsity. 
Such transformations aim  at representing 
a signal as a superposition of
elements, which are called `atoms' and are selected from a large set 
called `dictionary'. The superposition is said to be sparse if it involves a number of atoms  
 much smaller than the number of numerical values representing 
 the original signal. 

Given a dictionary, the problem of finding the sparsest 
approximation of a signal, up
to some acceptable error, is an NP-hard problem \cite{Nat95}.
In practice it is addressed  
by tractable methodologies known as Pursuit Strategies. 
Such methodologies can be  grouped 
for the most part in two broad categories. Namely, 
Basis Pursuit and Greedy Pursuit Strategies.
The Basis Pursuit (BP) approach endeavors to obtain a tractable
sparse solution by minimization of the 1-norm \cite{CDS01}.
Greedy algorithms seek for a sparse solution by stepwise
selection of dictionary's atoms.  When dealing with 
real data the latter  
are in general more convenient. 
From the seminal Matching Pursuit (MP) \cite{MZ93} and 
Orthogonal Matching Pursuit (OMP) \cite{PRK93} methods, a number 
of Greedy Pursuit Strategies have been developed 
to improve the process of sparsely representing single signals
\cite{Gri01, RNL02, ARNS04, ARN06, DTD06, NT09, NV10, EKB10, RNB13, 
 RNMB13, LRN16, RNA16}.  Due to complexity
issues and memory requirements, 
most of these techniques are to be   
 applied by segmenting the 
signal and approximating each segment independently of the others. 
Nonetheless, when the segments bear similarity with each other,  
for some applications it is convenient to look for the 
dictionary's atoms suitable to represent all the segments simultaneously. 
The greedy Pursuit Strategy which has been dedicated to 
 simultaneously approximate a set of signal is based on OMP 
 \cite{PRK93}  and 
 has been termed  Simultaneous Orthogonal Matching Pursuit  (SOMP)
 \cite{TGS06}. Since in this work we extend the Optimized Orthogonal Matching Pursuit method \cite{RNL02} to simultaneously approximate a set of signals, we term the new approach Simultaneous Optimized Orthogonal Matching Pursuit (SOOMP).

 The difference between  SOMP  and the  
 SOOMP approach introduced in this work is equivalent to the 
 difference between OMP  and   
 OOMP  methods, both for approximating single signals. 
 OOMP is stepwise optimal in the sense of minimizing at 
 each iteration the norm of the residual error. Whilst 
 OMP minimizes the norm of the error only with respect to the coefficients of the atomic superposition, OOMP  minimizes the norm of the error with respect to those coefficients  {\em{and}} the 
 selection of a new atom. 
 In the case of multiple signals SOOMP is designed to  
 minimize the 
 mean value of the error norm squared.  
 An additional advantage arises from the proposed 
 implementation. Based on adaptive biorthogonalization, the SOOMP 
 method  produces at each iteration the common dual 
 basis to the basis of selected atoms. This allows to calculate 
 the coefficients of the representation of each signal in the set 
 simply by computation of inner products. {{We implement the 
 previous SOMP method in an equivalent manner and 
  compare the two approaches 
 for the approximation of stereo music,  
 by selecting atoms 
 from a highly coherent trigonometric dictionary}}. 
 The practical relevance of the SOOMP approach is further 
 illustrated 
 by using it for compression of electrocardiogram (ECG) records. 

 An ECG signal represents a sequence of heartbeats which,  if  
 properly segmented and aligned, are suitable to be 
 simultaneously approximated.  This property  is 
  shown to benefit compression. Reliable comparison 
  with other compression techniques is made possible by 
  recuse to an adaptive quantization 
  procedure that facilitates to reconstruct the whole ECG record at the 
  required quality. The compression results are shown to significantly
  improve upon results produced by different transformation based approaches. 

  The paper is organized as follows: 
  Sec.~2  introduces the   
  problem and the mathematical notation. Sec.~3 establishes 
  the proposed SOOMP approach for simultaneous approximation 
 of a set of similar signals. {{Sec.~4 compares the SOOMP and 
 SOMP approaches for the simultaneous approximation  of 
 stereo music.}} Sec.~5 applies the proposed SOOMP method for 
 compressing digital ECG records and produces reliable 
 comparisons with previously reported results. 
 The conclusions are presented in Sec.~6. 

\section{Mathematical introduction of the problem}
\label{SOOMP}
In order to pose in mathematical terms the problem to be 
addressed we need to introduce 
  the notation used throughout the paper as well as
some preliminary background.

The sets of real, integer, and natural numbers are indicated by
$\R, \Z$, and  $\N$, respectively.
Boldface letters are used to indicate Euclidean vectors or matrices 
whilst standard mathematical fonts indicate components,
e.g., $\vf \in \R^N,\, N \in \N$ is a vector of components
$f(i),\, i=1,\ldots,N$ and  $\vC \in  \R^{Q \times k}$ is a matrix 
 of elements $C(i,j),\,i=1,\ldots,Q,\, j=1,\ldots,k$ which 
 when not leaving room for
ambiguity  will also be represented as $C(:,j), j=1,\ldots,k$.
A set of $Q$ signals of equal length $N$, 
to be simultaneously approximated
in a common subspace,
is represented
as a set of vectors $\{\vfq \in \R^{N},\,
q=1,\ldots,Q\}$. 
The inner product is indicated as $ \la  \cdot, \cdot \ra$, 
e.g. for $\vf\{1\} \in \R^{N}$ and  $\vf\{2\} \in \R^{N}$
$$\la  \vf\{1\}, \vf\{2\} \ra= \sum_{i=1}^{N} f\{1\}(i)f\{2\}(i).$$
The 2-norm induced by the inner product is 
denoted as $\|\cdot\|$, e.g. for $\vfq \in \R^{N}$
 $$\| \vfq \|= \sqrt{\la \vfq, \vfq\ra} = 
 \sqrt{\sum_{i=1}^{N} \left(f\{q\}(i)\right)^2}.$$

A set of $M$ vectors
$$\D=\left\{\vd_n \in \R^{N}\,; \|\vd_n\|=1 \right\}_{n=1}^M,$$
such that $\Spann\left(\D\right)=\R^{N}$ and $N < M$,
is called a redundant {\em{dictionary}} for $\R^N$ and its elements 
are called {\em{atoms}}.

In our context a signal $\vf \in \R^N$  is 
 assumed to be well approximated by an element, say $\vf^k$, 
 belonging to a finite dimensional subspace $V_k \subset \R^N$. 
 This assumption implies that, within a tolerance  $\rho$ 
 much larger than the numerical errors in the calculations, 
 $\vf^k \in V_k$ is accepted to 
 be a good approximation of $\vf \in \R^N$ if
$\|\vf - \vf^k \| < \rho.$ 
 Examples of signals fulfilling this definition are, amongst others, 
 audio signals, and electrocardiograms. These are 
 all signals with acceptable approximations 
 which,  without  affecting their informational content,   
 do not necessarily produce a highly accurate 
 point-wise reproduction of the signals. These type 
 of signals are suitable for lossy compression. 

 Since this work concerns approximation of 
 similar signals we need to make an assumption on the signals that 
 will be considered. We say that a finite set of $Q$ signals
$\{\vfq \in \R^N\}_{q=1}^Q$ are similar if they can be 
 well approximated in a subspace $V_k$ of dimension $k$, 
with $k$ significantly smaller than $N$.  
This is equivalent to assuming that 
there exists a common basis $\{\vd_{\ell_n}\}_{n=1}^k$
for $V_k$ such that each signal $\vfq$ is approximated as
$$\vft\{q\}= \sum_{n=1}^k c\cq(n) \vd_{\ell_n},\quad q=1,\ldots,Q.$$
The quality of the approximated set will be assessed in mean value
$${\overline{{\cal{E}}}} = \sum_{q=1}^Q p(q) \| \vfq - \vft\{q\}\|^2,$$
where $p(q)\ge 0$  with $\sum_{q=1}^Q p(q)=1$.

\section{Strategy for simultaneous approximation of a set of  
signals}
\label{SOOMP}
Given a set of similar signals 
$\{\vfq \in \R^{N}\}_{q=1}^Q$
and a
dictionary, 
the aim  is to simultaneously 
approximate all the signals in the set $\{\vfq \in \R^{N}\}_{q=1}^Q$ within a common subspace $V_k=\Spann\left(\{\vd_{\ell_n}\}_{n=1}^k\right)$. 
In other words,   
each signal $\vfq \in \R^{N}$ is to be approximated as a 
$k$-term atomic superposition 
\be
\label{atoq}
\vfq^k= \sum_{n=1}^{k}
 c\cq(n) \vd_{\ell_n},
\quad q=1,\ldots,Q,
\ee
where the atoms $\vd_{\ell_n},\, n=1,\ldots,k$ in \eqref{atoq} are 
 selected from the given dictionary according to the criterion of 
optimality that will be established by Proposition \ref{P1}
in the next subsection.
Let us suppose for the moment that these atoms 
are known. Assigning a weight  
$p(q)\ge 0$ to the signal $\vfq$, with 
$\sum_{q=1}^Q p(q)=1$, 
the coefficients $\vc\cq \in \R^{k}$ in \eqref{atoq} are required
to minimize the mean value of the square norm of the errors in the 
approximation of
the set of signals, i.e. 
\be
\vc\cq,\ldots,\vc\{Q\}= 
\operatorname*{arg\,min}_{\substack
\vc'\cq,\ldots,\vc'\{Q\}}
\sum_{q=1}^Q  p(q) \|\vfq - \sum_{n=1}^{k}
 c'\cq(n) \vd_{\ell_n}\|^2.
\ee
Since $p(q)\ge 0$ the above minimization is 
equivalent to finding the components 
$c\cq(n),\,n=1,\ldots,k$ of each vector 
$\vc\cq$ such that
\be
c\cq(1),\ldots, c\cq(n)=
\operatorname*{arg\,min}_{\substack 
c'\cq(1),\ldots,c'\cq(n)} \|\vfq - \sum_{n=1}^{k}
 c'\cq(n) \vd_{\ell_n}\|^2\,\quad q=1\ldots,Q.
\ee
Accordingly, 
the minimization with respect to the coefficients in 
\eqref{atoq} can be implemented by adaptive biorthogonalization 
\cite{LRN07}, as proposed 
within the 
OOMP algorithm for a single signal \cite{RNL02}, 
\be
c\cq(n)= \la \vbeta_n^k , \vfq\ra, \quad q=1,\ldots,Q,
\ee
with vectors $\vbeta_n^k$ 
calculated as will be described in the next section.

The selection of the atoms $\vd_{\ell_n},\,n=1\ldots,k$ 
in the decomposition
\eqref{atoq} such that
 $$\sum_{q=1}^Q p(q) \|\vfq - \sum_{n=1}^{k} c\cq(n) \vd_{\ell_n}\|^2\,\,\,\text{is minimized}$$ poses an intractable 
 problem (for a dictionary of $M$ atoms there
are $\frac{M!}{(M-k)!k!}$ possibilities to be checked). 
We address the 
selection in a tractable manner by extending 
the OOMP strategy to simultaneously approximate 
 a set of similar 
signals. The extended strategy is refereed to as
SOOMP (Simultaneous OOMP).
\subsection{SOOMP algorithm}
The algorithm is initialized by setting:
$\vrq^0=\vfq$, $ \vfq^0=0$, $\Gamma= \emptyset$
and $k=0$.
 The first atom
 is selected as the one corresponding to the index
$\ell_{1}$ such that
\be
\label{msoomp1}
\ell_{1}=\operatorname*{arg\,max}_{n=1,\ldots,M}
  \sum_{q=1}^Q p(q) \left |\la \vd_n,  \vrq^{0} \ra \right|^2.
\ee
This first atom is used to
assign $\vW_1= \vbeta_1=\vd_{\ell_{1}}$,
calculate $\vrq^{1}= \vfq - \vd_{\ell_{1}} 
\la \vd_{\ell_{1}}, \vfq\ra$ and iterate
 as prescribed below.

\begin{itemize}
\item[1)] Upgrade the set $\Gamma \leftarrow  \Gamma \cup 
\ell_{k+1}$, increase $k \leftarrow k +1$, and
 select the index of a new atom for the approximation as
\be
\label{msoomp}
\ell_{k+1}=\operatorname*{arg\,max}_{\substack{n=1,\ldots,M\\ n\notin \Gamma}}
 \sum_{q=1}^Q p(q) \frac{\left|\la \vd_n,\vrq^{k} \ra\right|^2}{1 - \sum_{i=1}^{k}
|\la \vd_n ,\vWt_i\ra|^2},
 \quad \text{with} \quad \vWt_i= \frac{\vW_i}{\|\vW_i\|}.
\ee
\item[2)]
Compute the corresponding new vector $\vW_{k+1}$ as
\be
\begin{split}
\label{GS}
\vW_{k+1}= \vd_{\ell_{k+1}} - \sum_{i=1}^{k} \frac{\vW_i}
{\|\vW_i\|^2} \la \vW_i, \vd_{\ell_{k+1}}\ra,
\end{split}
\ee
including for numerical accuracy  the
re-orthogonalization step:
\be
\label{RGS}
\vW_{k+1} \leftarrow \vW_{k+1}- \sum_{i=1}^{k} \frac{\vW_{i}}{\|\vW_i\|^2}
\la \vW_{i} , \vW_{k+1}\ra.
\ee
\item[3)]
 Upgrade vectors $\vbeta_n^{k}$ as
\be
\label{adapt}
\vbeta_{k+1}^{k+1}= \frac{\vW_{k+1}}{\|\vW_{k+1}\|^2}, \quad
\vbeta_{n}^{k+1}= \vbeta_{n}^{k} - \vbeta_{k+1}^{k+1}\la \vd_{\ell_{k+1}}, \vbeta_{n}^{k}\ra,\quad n=1,\ldots,k.
\ee
\item [4)] Update $\vrq^{k}$ as
\ben
\vrq^{k+1} &=& \vrq^{k} - \la \vW_{k+1}, \vfq \ra  \frac{\vW_{k+1}}{\|\vW_{k+1}\|^2}.
\een
\item[5)] If the stopping criterion is met finish the 
iterations. Otherwise repeat steps 1) - 5).
\end{itemize}
{{{\bf{Note}}: given a tolerance $\rho$, 
as stopping criterion we set:
\be
\label{sc}
\sum_{q=1}^Q p(q) \|\vrq^{k+1}\|^2 < \rho
\ee
or 
\be
\label{sc2}
\sum_{q=1}^Q p(q) \|\vrq^{k+1}\| < \rho,
\ee
depending on convenience for the particular application.}}\\
Once the iterations have finished calculate the coefficients for the decomposition \eqref{atoq} as
$$c\cq(n)=\la \vbeta_n^k,\vfq\ra,\quad n=1,\ldots,k,\, 
q=1,\ldots,Q.$$
For $q=1,\ldots,Q$ calculate the final approximation of each signal $\vfq$ as
$$\vfq^k = \vfq -\vrq^k.$$

\begin{remark}
\label{R1}
The set of vectors $\vbeta_n^{k},\,n=1,\ldots,k,\,
q=1,\ldots,Q$ as given in 
\eqref{adapt} fulfills that
$$\vfq^k=  \sum_{n=1}^{k}
 \la \beta_n^k, \vfq\ra  \vd_{\ell_n}= \hat{P}_{V_k} \vfq,\quad q=1,\ldots,Q,$$
where $ \hat{P}_{V_k} \vfq$ is the orthogonal projector of 
$\vfq$ onto 
$V_k=\Spann\{\vd_{\ell_n}\}_{n=1}^k$. Please find the proof in \cite{RNL02}, or 
as a particular case of the more general proof 
in \cite{LRN07}.
\end{remark}

\begin{proposition}
\label{P1}
The recursive selection of the indices 
$\ell_1,\ldots,\ell_{k}$, as proposed in \eqref{msoomp},
is stepwise optimal. It  
minimizes, at each iteration, the mean of the square 
distance between 
the set of signals $\vfq,\,q=1,\ldots,Q$ 
and their corresponding approximations
 $\vfq^k,\,q=1,\ldots,Q$.  
\end{proposition}
\begin{proof}
 For $k=0$ it is clear that $\ell_1$ selected as in
 \eqref{msoomp1} minimizes the mean  of the square distance
$\overline{\E^1}$ as given by
$$\overline{\E^1}=\sum_{q=1}^Q p(q) \|\vfq -  \vfq^1\|^2=\sum_{q=1}^Q p(q) (\| \vfq\|^2-
|\la\vd_{\ell_1},\vfq \ra |^2).$$
Let us assume that the indices $\ell_1,\ldots,\ell_{k}$
selected as proposed in \eqref{msoomp} minimize, in 
the specified stepwise sense, the mean square
distance 
$$\overline{\E^k}=\sum_{q=1}^Q p(q) \|\vfq -  \vfq^k \|^2.$$
We shall prove by induction that if the atoms
 $\vd_{\ell_1},\ldots,\vd_{\ell_{k}}$ are fixed, at iteration $k+1$ the atom $\vd_{\ell_{k+1}}$ selected as in \eqref{msoomp}
 minimizes $\overline{\E^{k+1}}$.  
The proof stems from the
fact that at iteration $k$ the approximation $\vfq^k$ of
each signal $\vfq$ is the orthogonal projection of
$\vfq$ onto the subspace $V_k=\Spann\{\vd_{\ell_n}\}_{n=1}^k$
(c.f. Remark \ref{R1}). 

Consider that $V_k$ is augmented by one element, say
$\vd_{\ell_{k+1}} \notin V_k$, so that
$V_{k+1}=V_k \oplus \vd_{\ell_{k+1}}$, where $\oplus$ indicates  direct sum. The orthogonal projection of each signal $\vfq,\,q=1,\ldots,Q$ onto $V_{k+1}$ can be expressed as
$$\vfq^{k+1}= \hat{P}_{V_{k+1}} \vfq =\hat{P}_{V_k} \vfq + 
\frac{\vW_{k+1}}{\|\vW_{k+1}\|^2}
\la \vW_{k+1}, \vfq\ra\quad \text{with} \quad \vW_{k+1}=
\vd_{\ell_{k+1}} - \hat{P}_{V_k}  \vd_{\ell_{k+1}}.$$
Thus
\ben
\|\vfq - \vfq^{k+1}\|^2 &=& \|\vfq - \hat{P}_{V_{k+1}} \vfq\|^2\nonumber\\
&=&\|\vfq - \hat{P}_{V_{k}} \vfq - \frac{\vW_{k+1}}{\|\vW_{k+1}\|^2}  \la \vW_{k+1}, \vfq\ra\|^2\nonumber\\
&=& \|\vfq - \hat{P}_{V_{k}} \vfq\|^2 - \frac{|\la \vW_{k+1}, \vfq\ra|^2}{\|\vW_{k+1}\|^2}. \nonumber
\een
Since $\|\vfq - \hat{P}_{V_{k}} \vfq\|^2$ is optimized  and
fixed  at iteration $k$, it is true that at iteration
$k+1$ the index of the atom  which 
 minimizes $\overline{\E^{k+1}}$ fulfils
\ben
\label{sel2}
\ell_{k+1}&=&
\operatorname*{arg\,max}_{\substack{n=1,\ldots,M\\ n\notin \Gamma}} \sum_{q=1}^Q p(q) \frac{|\la \vW_{k+1}, \vfq\ra|^2}{\|\vW_{k+1}\|^2}\nonumber\\
&=&
\operatorname*{arg\,max}_{\substack{n=1,\ldots,M\\ n\notin \Gamma}}
 \sum_{q=1}^Q p(q) \frac{\left|\la \vd_{\ell_{n}} - \hat{P}_{V_{k}}\vd_{\ell_{n}},\vfq \ra \right|^2}{\|\vd_{\ell_{n}}- \hat{P}_{V_{k}} \vd_{\ell_{n}}\|^2}.
\een
The proof is concluded using the self-adjoint properties of
$\hat{P}_{V_{k}}$ to write:
\ben
\label{equi}
\left |\la \vd_{\ell_{n}} - \hat{P}_{V_{k}}\vd_{\ell_{n}}, \vfq\ra \right|
& =& \left |\la \vd_{\ell_{n}}, \vfq \ra - \la \hat{P}_{V_{k}} \vd_{\ell_{n}} , \vfq\ra \right|
\nonumber\\
& =& \left | \la \vd_{\ell_{n}}, \vfq -  \hat{P}_{V_{k}} \vfq\ra \right|
\nonumber\\
&=& \left|\la \vd_{\ell_{n}}, \vrq^k \ra \right|.
\een
Moreover, since all atoms are normalized and the set 
$\{\vWt_i\}_{i=1}^k$ is an orthonormal basis for $V_k$ 
we have
$$\hat{P}_{V_{k}} \vd_{\ell_{n}}= \sum_{i=1}^k  \vWt_i \la \vWt_i, \vd_{\ell_{n}}\ra,$$  so that
\be
\|\vd_{\ell_{n}}- \hat{P}_{V_{k}} \vd_{\ell_{n}}\|^2=
1-\|\hat{P}_{V_{k}} \vd_{\ell_{n}}\|^2=
1 - \sum_{i=1}^k |\la \vWt_i, \vd_{\ell_{n}}\ra|^2 \nonumber,
\ee
which shows the equivalence between
 \eqref{sel2} and \eqref{msoomp}.
\EndProof

\begin{corollary}
The selection criterion \eqref{msoomp} guaranties that, 	
if for $n=1,\ldots,k$ the right hand side of \eqref{msoomp} 
is not zero, 
the selected elements $\vd_{\ell_{n}},\,n=1,\ldots,k$ are 
linearly independent.
\end{corollary}

\Proof For $k=1$ the single 
atom $\vd_{\ell_{1}}$ is linearly independent. Let us assume that 
the first $k$ selected atoms $\vd_{\ell_{n}},\,n=1,\ldots,k$ 
 are linearly independent and prove that then the newly selected 
atom $\vd_{\ell_{k+1}}$ is also linearly independent. The proof is 
	achieved by contradiction. Indeed, if $\vd_{\ell_{k+1}}$ 
	is linearly dependent then  
	$\vd_{\ell_{k+1}}= \sum_{n=1}^k a_n \vd_{\ell_{n}}$, for 
	some scalers $a_n,\,n=1,\ldots,k$, so that $\hat{P}_{V_{k}} \vd_{\ell_{k+1}}= \sum_{n=1}^k a_n \hat{P}_{V_{k}}  \vd_{\ell_{n}}= 
	\sum_{n=1}^k a_n \vd_{\ell_{n}} = \vd_{\ell_{k+1}}$ and 
	from \eqref{equi} we gather 
	that the right hand side of \eqref{msoomp} is zero. This 
	contradiction leads to conclude that the selected elements $\vd_{\ell_{n}},\,n=1,\ldots,k$ by criterion \eqref{msoomp} 
	are linearly independent.
\EndProof	
\begin{remark} If the dictionary $\D=\left\{\vd_n \in \R^{N}\,; \|\vd_n\|=1 \right\}_{n=1}^M$ is complete or over-complete, i.e., 
	$\dim\left(\Spann\left(\D\right)\right)=N$, then 
	 by selecting $N$ atoms the method can reconstruct 
	 the exact signals  
	 $\vfq \in \R^{N},\,q=1,\ldots,Q.$ Otherwise if 
	 $\dim\left(\Spann\left(\D\right)\right) =S<N$
	 the selected atoms provide  the orthogonal projections 
	 $\hat{P}_{V_S} \vfq,\, q=1,\ldots,Q$, where 
	 $V_S= \Spann \left(\D\right)$. It  should be stressed, 
	 though, that with a suitable dictionary all the 
	 signals $\vfq, q=1,\ldots,Q$
	 are expected to be well approximated in a 
	 subspace $V_k$ with $k$ significantly smaller 
	 than $N$. Otherwise the representation
	 would not qualify to be sparse. 
\end{remark}

\begin{remark} The complexity of the SOOMP algorithm, at
       each iteration, is $O(N M Q)$.
       For equal weights 
$p(q)=\frac{1}{Q}, \,q=1,\ldots,Q$ the difference between the complexity of the SOOMP selection 
	criterion \eqref{sel2} and the SOMP one 
	\cite{TGS06} is the denominator 
	in the right hand side of \eqref{sel2}. 
	This introduces extra computations of 
	complexity $O(N M)$ 
	so that the order of complexity of both 
	algorithms is equivalent. However, 
	as will be illustrated in the next section, 
	due to the reduction in the number of 
	selected atoms, and hence the number of iterations,  
	the approximation may run faster if 
	SOOMP rather than SOMP is used.
\end{remark}
\section{Numerical Example} 
We illustrate here the algorithm's implementation by simultaneously 
approximating  stereophonic music. This type of 
music, commonly called stereo music, 
is usually produced by using two independent audio channels
$\vf\{1\}\in \R^N$ and  $\vf\{2\}\in \R^N$.
The example is dedicated to showing 
the sparsity 
obtained with the proposed SOOMP method and the precursor SOMP one, 
when simultaneously approximating both channels  
 using a highly coherent trigonometric dictionary.

Since the signals
structure varies with time, 
approximations with trigonometric dictionaries are carried out on a
partition of the signal. In this example the partition
consists of disjoint segments, called frames, which are assumed to be all of the same size $\Nq$. The signal representation is
realized by independent approximation of each frame.

As shown in \cite{LRN16,RNA16}, 
for music representation the combination of
a Redundant Discrete Cosine Dictionary  $\Dc$
and a Redundant Discrete Sine Dictionary  $\Ds$, 
defined below, renders higher sparsity than when using 
pure $\Dc$ or $\Ds$ dictionaries of the same redundancy 
as the combination $\D= \Dc \cup \Ds$.
\begin{itemize}
\item
$\Dc=\{{w^{\rc}(n)}
 \cos{\frac{{\pi(2i-1)(n-1)}}{2M}},i=1,\ldots,\Nq\}_{n=1}^{M}.$
\item
$\Ds=\{{w^{\rs}(n)}\sin{\frac{{\pi(2i-1)(n)}}{2M}},i=1,\ldots,\Nq\}_{n=1}^{M},$
\end{itemize}
where $w^{\rc}(n)$ and $w^{\rs}(n),\, n=1,\ldots,M$
are normalization factors.

The signal representation is realized by independent 
approximation of each frame. The 
atoms $\vd_{\ell_n^i}, \,n=1\ldots,k_i$ 
are selected for simultaneously representing both 
channels $\vf_i\{1\}\in \R^{\Nq}$
and $\vf_i\{2\}\in \R^{\Nq}$ in the $i$th frame, for $i=1,\dots,I$, with 
$I$ the integral part of $N/\Nq$.  
In the numerical simulations the number of atoms 
in each dictionary $\Dc$ and  $\Ds$ is $2\Nq$ so that 
the total number of atoms in $\D= \Dc \cup \Ds $ is $4\Nq$.

As a metric of approximation
quality we use the standard Signal to Noise Ratio
($\SNR$) calculate as
$$\SNR=10 \log_{10}\left(\frac{\sum_{q=1}^2 \frac{1}{2}\|\vf\{q\}\|^2}{\sum_{q=1}^2 \frac{1}{2}\|\vf\{q\} - \vfa\{q\}\|^2}\right),$$
where $\vfa\{q\} \in \R^N,\, q=1,2$ are the approximations of the
channels $\vf\{q\} \in \R^N,\, q=1,2$. Since the
frames are disjoint the approximation of each channel
$\vfa\{q\}$ is obtained by the concatenation of the approximations
$\vf_i\{q\}^{k_i}$ 
of the corresponding frames i.e., $\vfa\{q\} = \Join_{i=1}^I 
\vf_i\{q\}^{k_i}, \, q=1,2$, where $\Join$ indicates the 
concatenation operation. The numbers $k_i,\,i=1,\ldots,I$ 
of atoms for approximating 
each $i$-frame are decided to meet the condition:
\be
\label{rhosnr}
\sum_{q=1}^2 \frac{1}{2}\|\vf_i\{q\} - \vf_i\{q\}^{k_i}\|^2< 
\rho_i \quad i=1,\ldots,I.
\ee
On defining
$$\snr_i= 10 \log_{10}\left(\frac{\sum_{q=1}^2 \frac{1}{2}\|\vf_i\{q\}\|^2}{\sum_{q=1}^2 \frac{1}{2}\|\vf_i\{q\} - \vf_i\{q\}^{k_i}\|^2}\right), \quad i=1,\ldots,I$$
the parameters $\rho_i,\,i=1,\ldots,I$ in \eqref{rhosnr} 
are determined as 
$$\rho_i= 10^{(-\frac{\snr_0}{10})}\sum_{q=1}^2\frac{1}{2}\|\vf_i\{q\}\|^2,\quad i=1,\ldots,I,$$
where $\snr_0$ is fixed to the same value for 
each frame in oder to achieve the expected  approximation 
quality.

The metric of sparsity is 
considered to account for the  number of 
elements in the approximation of the whole signals. 
For this we define the Sparsity Ratio (SR) as
$$\text{SR}= \frac{2 N}{K},$$ where $K=\sum_{i=1}^I k_i$ 
is the total number
of atoms in the signal representation and $N$ the number of
samples in each of the channels. Thus, 
a large  value of
$\SR$ indicates a high level of sparsity.

The numerical example is realized using four stereo clips of 
melodic music: 1. Classic Orchestra.  2. Classic Guitar. 
3. Chopin Piano. 4. Piazzolla Tango.
All four clips are of the same length, $N=256000$ samples
in each channel (5.8 secs). 
The approximation is carried out 
on frames of length $L=1024$ and 
for qualities corresponding to $\SNR=20$dB, $\SNR=25$dB, and 
$\SNR=30$dB. The improvements in the  values of $\SR$ 
obtained with SOOMP, with respect to those obtained with SOMP, 
are noticeable from the comparison of the 3rd and 4th columns 
in Table \ref{Tab_mu} for all  four clips and 
all approximation qualities. The 5th column displays the gain over
SOMP yielded by SOOMP. The 6th and 8th columns show the approximation times (the values are the average of five independent runs 
with MATLAB in a Laptop Core i7-1165G7). 
The 7th and 9th columns give the corresponding standard deviations. 

\begin{table}[!ht]
{{	
\begin{center}	
\begin{tabular}{| l|c|c|c|c|c|c|c|c|}
\hline
	Clip  &SNR&SOMP(1) &SOOMP(2)&Gain & Time (1) & std &  Time (2) &std\\ \hline \hline
1. &20dB& 43.3 & 48.4 &11.8\% & 41.64  s& 0.08 & 37.25 s& 0.07\\ \hline
2. &20dB& 119.4 & 135.7& 13.6\% & 15.67 s & 0.11 & 13.88 s& 0.07\\ \hline 
3. &20dB& 67.3  &74.5 &10.7\% &26.32 s&0.38 & 23.91 s & 0.22\\ \hline
4. &20dB& 94.9 & 106.9&12.6\% &18.42 s & 0.08 & 16.41 s & 0.02\\ \hline \hline
1. &25dB& 28.0& 31.3&11.8\% &66.81 s &0.06& 59.61 s&0.05\\ \hline
2. &25dB& 78.8& 90.7 &15.1\% & 24.13 s & 0.13 & 20.93 s &0.23 \\ \hline
3. &25dB& 45.7& 51.3 &12.2\% & 39.19 s & 0.09& 35.20 s & 0.29\\ \hline
4. &25dB&69.3& 79.0 &14.0\%&25.53 s & 0.04 & 22.34 s & 0.03 \\ \hline \hline
1. &30dB& 17.9 & 19.8 &10.6\% &113.01 s &0.21& 102.17 s& 0.22\\ \hline
2. &30dB& 53.9 & 61.8 &14.7\% & 35.79 s & 0.27& 31.02 s & 0.08 \\ \hline
3. &30dB& 32.3 & 36.3 &12.4\% & 57.01 s&0.07& 50.91 s& 0.32\\ \hline 
4. &30dB& 54.4 & 62.1 &14.5\% & 33.06 s & 0.29& 28.73 s& 0.06 \\ \hline 
\end{tabular}
\caption{\small{Comparison of sparsity (SR values) for
	approximations of four clips of music up to 
	the same SNR (20, 25 and 30dB).
       The 4th column gives the 
	gain in SR achieved by SOOMP (2) over SOMP (1). 
	The 6th and 8th columns give the
	approximation times in secs. (average of five independent runs).
	The 7th and 9th columns are the corresponding standard 
	deviations.}}
\label{Tab_mu}
\end{center}
	}}
\end{table}

The left graphs in Fig.\ref{frames} 
show 2000 samples in channel 
1 of the original chips as well as the corresponding 
approximation up to 25dB.  The 
right graphs have the same description but correspond to  
the 2000 samples in channel 2. As illustrated by the 
graphs in Fig.\ref{frames}, SNR=25dB produces already 
very good pointwise approximation of the signals.

\begin{figure}[ht!]
\centering
\includegraphics[width=6cm]{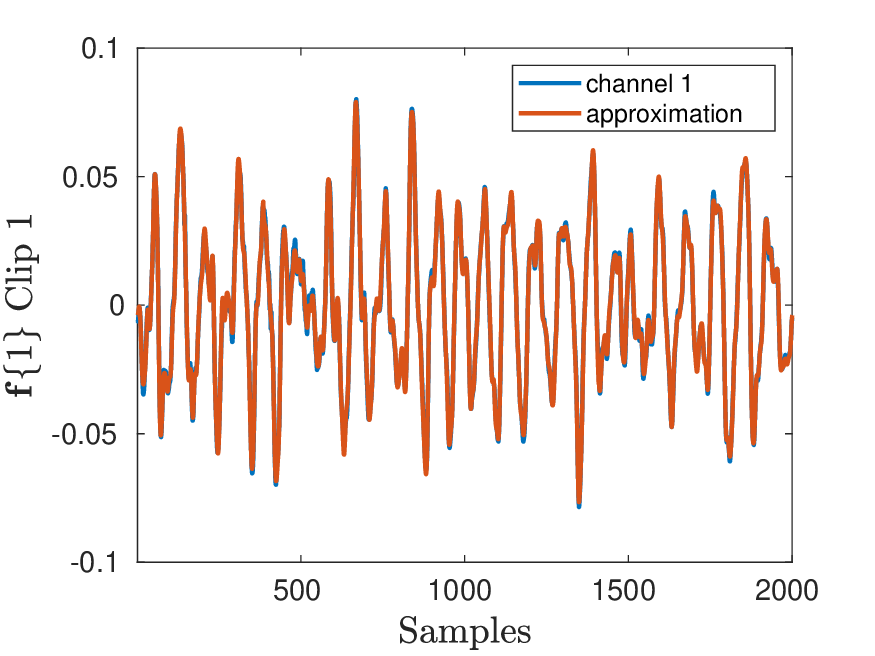}
\includegraphics[width=6cm]{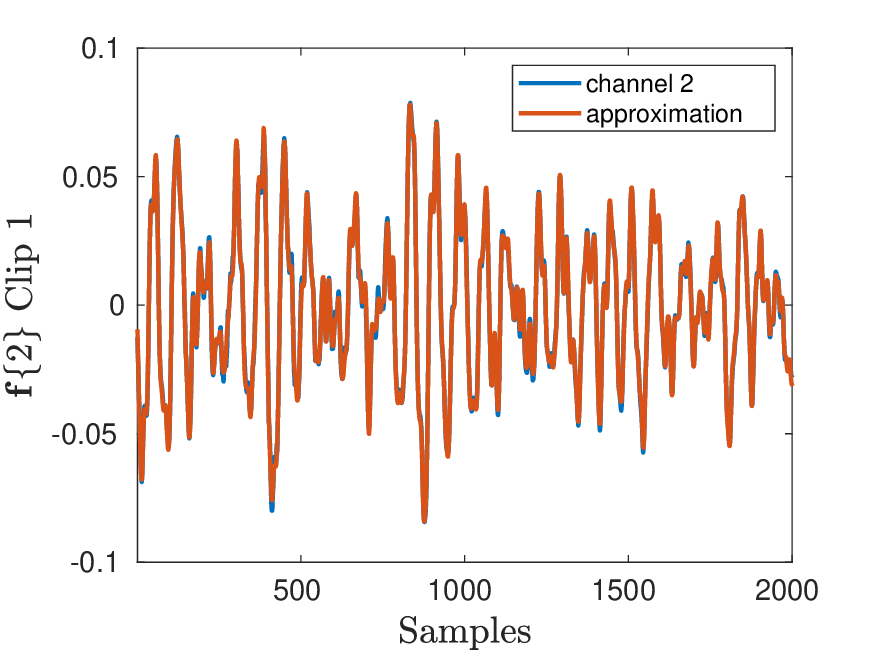}\\
\includegraphics[width=6cm]{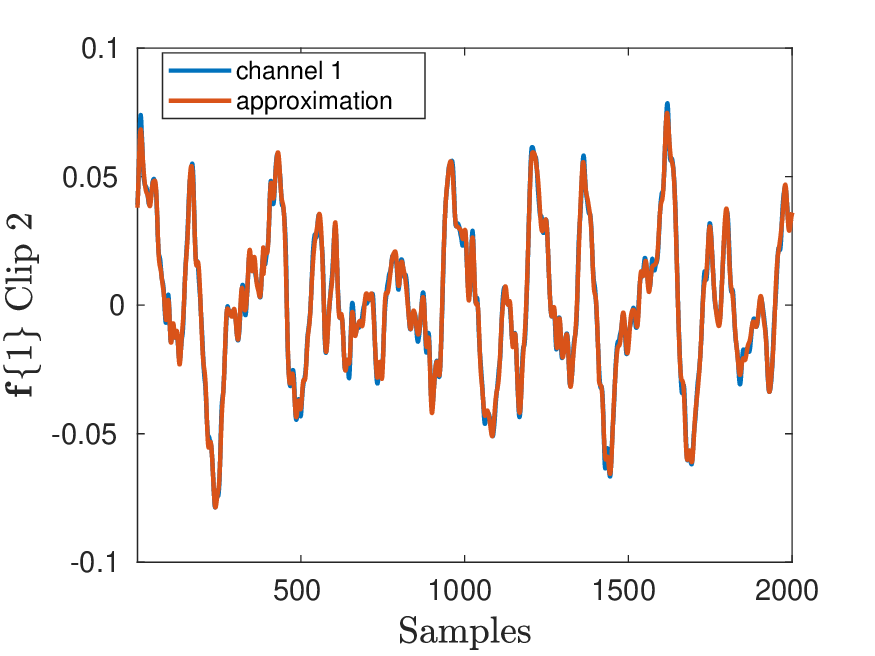}
\includegraphics[width=6cm]{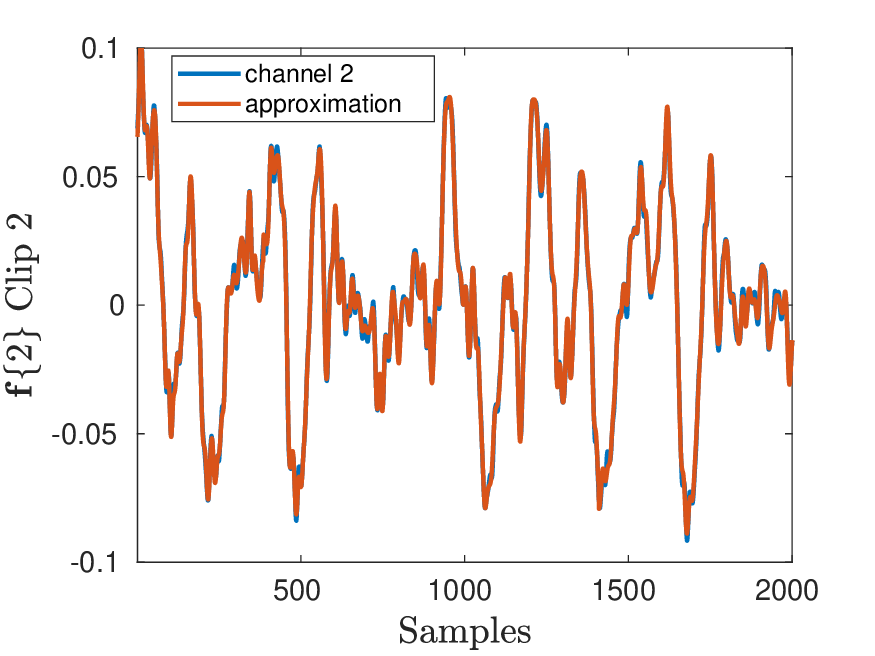}\\
\includegraphics[width=6cm]{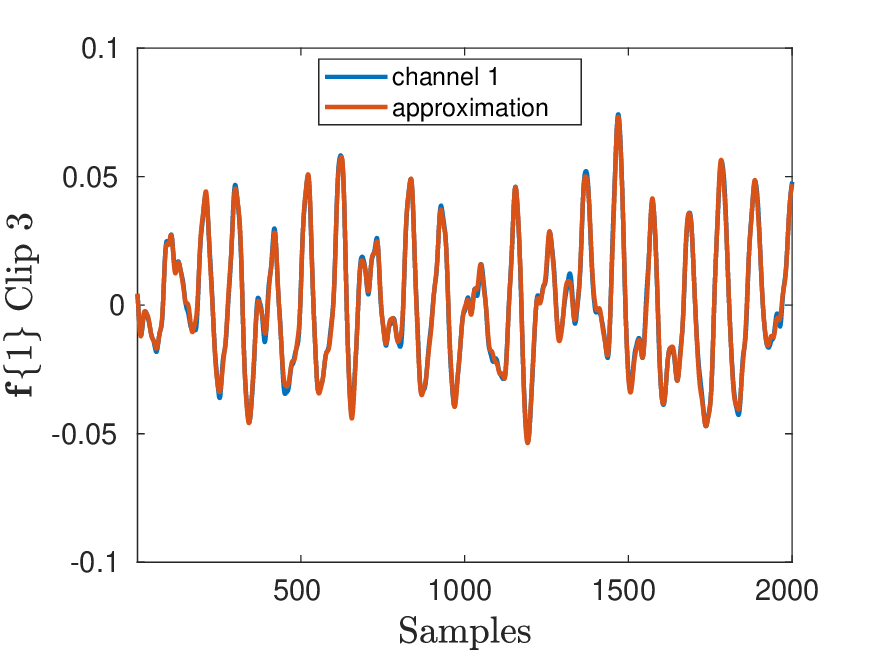}
\includegraphics[width=6cm]{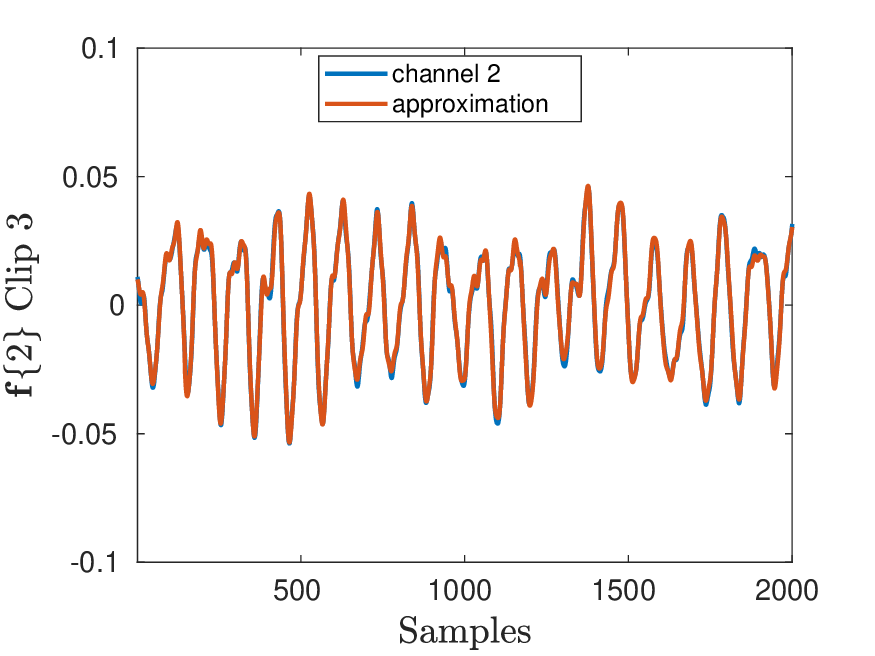}\\
\includegraphics[width=6cm]{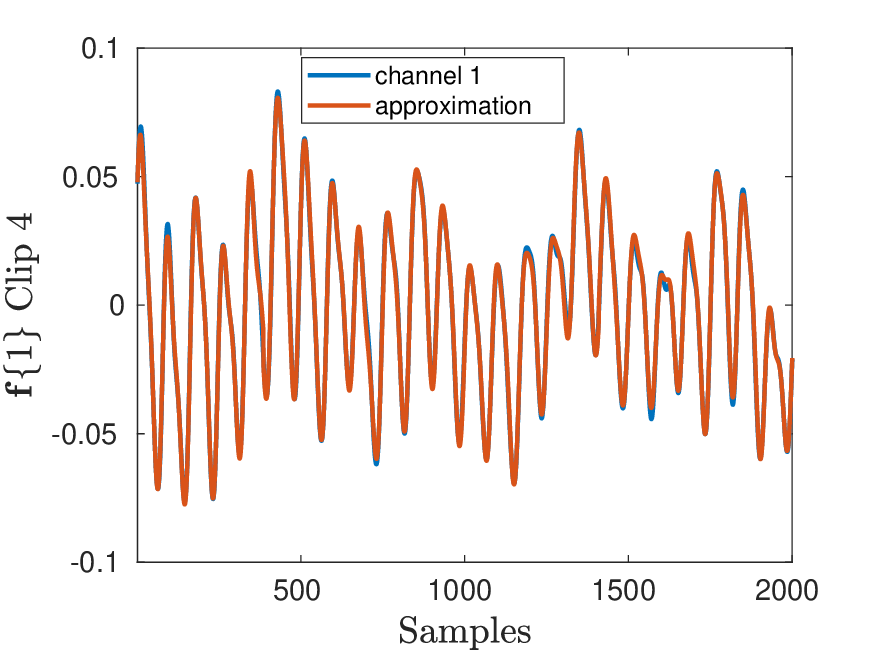}
\includegraphics[width=6cm]{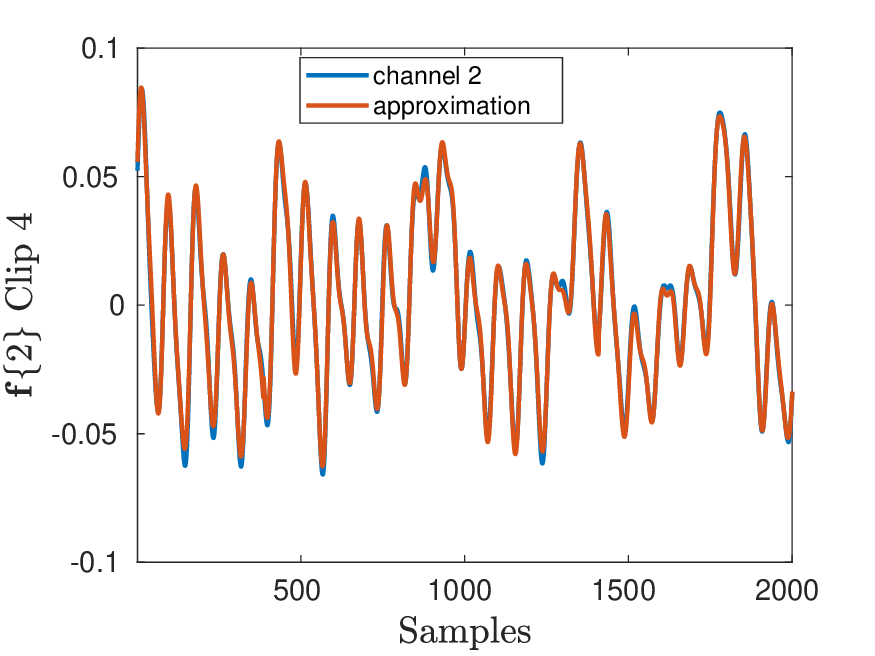}\\
\caption{2000 samples in the Chips 1 - 4 (blue lines) 
and the corresponding approximations (red lines) 
up to SNR=25dB.  The graphs on the left 
correspond to one of the channels and the graphs on the right 
to the other. \label{frames}}
\end{figure}
\newpage
\section{Application to compression of ECG records}
A digital ECG signal represents a
 sequence of heartbeats. In a typical record each heartbeat is 
 characterized by a 
combination of three graphical deflections, known as QRS complex, 
and two lateral and less visually noticeable P and T waves.
A short segment of a typical ECG record is illustrated in 
Fig.\ref{peaks}. 

\begin{figure}[ht!]
\begin{center}
       \includegraphics[width=80mm]{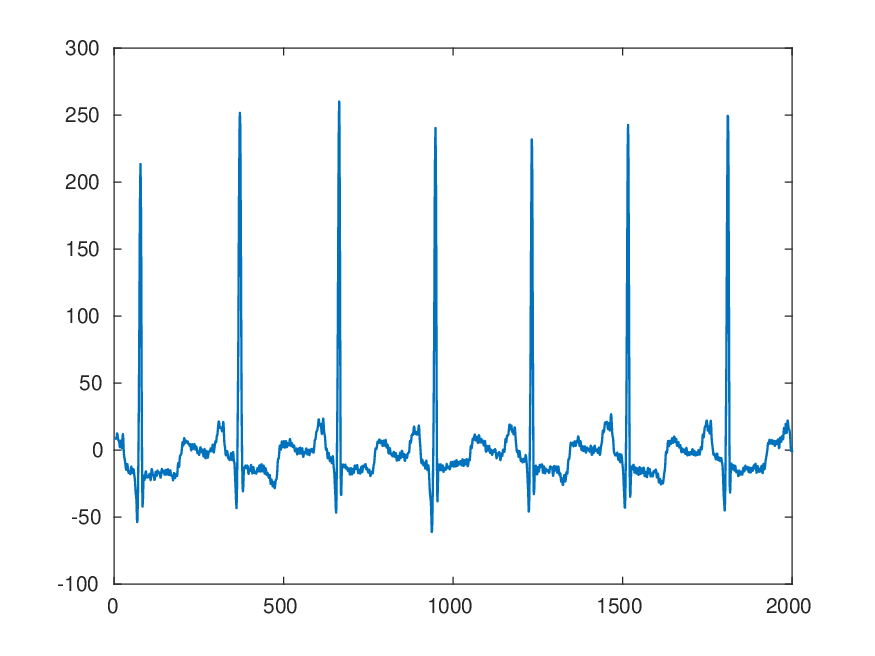}
\end{center}
\caption{A short segment of an ECG record \label{peaks}}
\end{figure}

In order to simultaneously 
approximate all the beats in a record we need to 
segment and align the beats to meet the requirement of being
similar. The procedure is discussed in the next 
subsection.

\subsection{Segmentation and alignment of heartbeats}
The QRS complex is segmented once the  
central R peak is detected. This can be effectively done  
by the Pan Tompkins method 
\cite{PT85}. In our numerical examples we  
 use the 
off-the-shelf MATLAB implementation of this algorithm \cite{Sed14}.
Since the distance between peaks in a record is not uniform, 
 the length of the segmented beats 
 should be  passed to the 
 decoder. The segmented peaks are placed in 
 arrays $\vfq,\,q=1,\ldots,Q$
 of equal length $\Nq$ by padding with zeros.
 Fig.\ref{alig} illustrates the resulting configuration with 
 80 heartbeats.{{
Fig.~\ref{2Dbeats} shows the two dimensional image of the
segmented and aligned heartbeats corresponding to
records 111 and 100 in the MIT-BIH Arrhythmia database \cite{MITDB}.

\begin{figure}[ht!]
\begin{center}
\includegraphics[width=9cm]{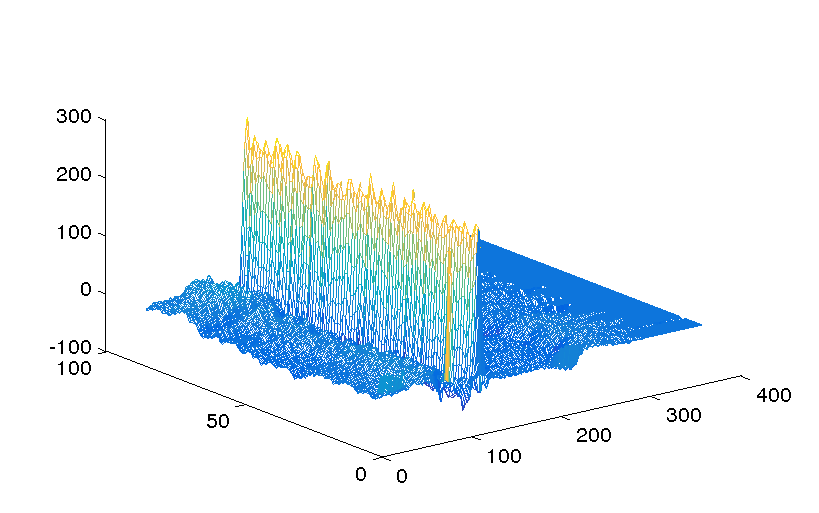}
\end{center}
\caption{Configuration resulting by segmentation and alignment of
 80 heartbeats  for illustration purposes.}
\label{alig}
\end{figure}

\begin{figure}[ht!]
\begin{center}
\includegraphics[width=6.45cm, height=5cm]{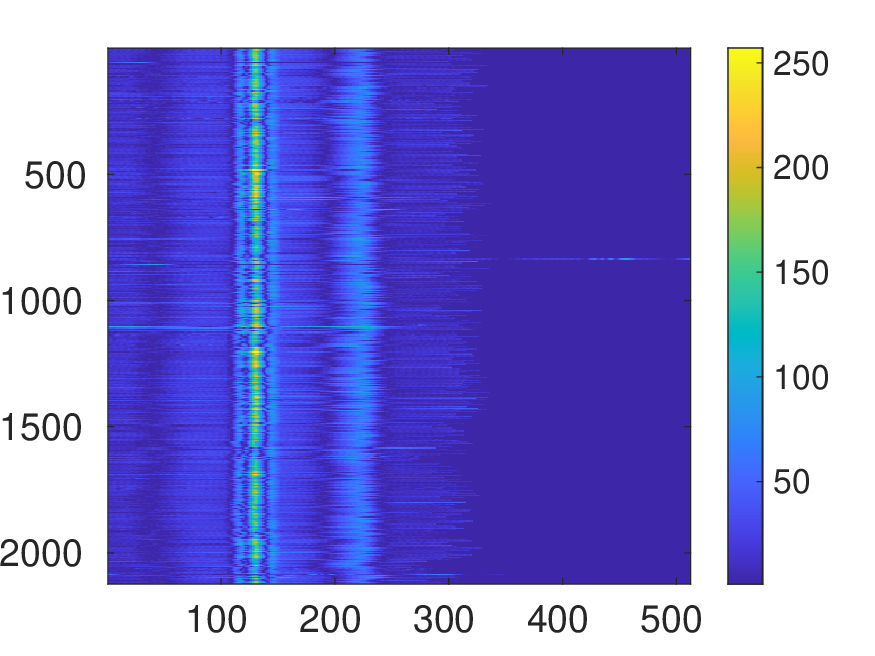}
\includegraphics[width=6cm, height=5cm]{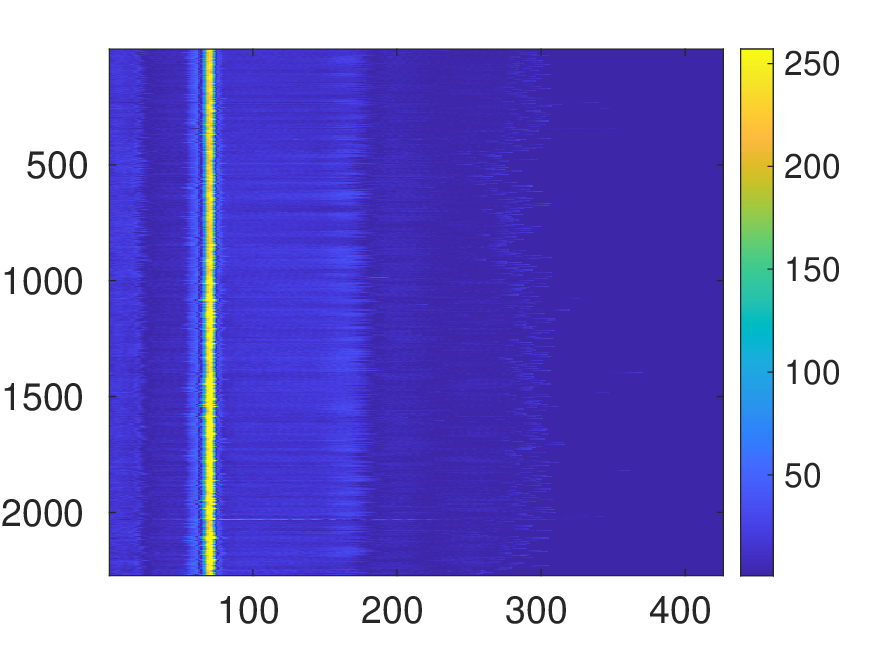}
\end{center}
\caption{Images of the magnitude of the aligned heartbeats 
	in records 111 (left graph) and 100 (right graph).} 
\label{2Dbeats}
\end{figure}
}}
For simultaneously approximating heartbeats we use a 
wavelet dictionary. Given 
 a partition $x_i,\, i=1,\ldots,N$ of the interval $[c,d]$ the 
  dictionary is constructed as
 follows \cite{RNC19, CRN20}. 
\be
\label{dic}
\D=\dicv_{0}\cup \dicw_{0} \cup \dicw_{1} \cup \dicw_{2} \cup \dicw_{3}
  \cup \dicw_{4},
\ee
with
\be
\dicv_0=
\{\phi(x_i-\frac{k}{2})|_{[c,d]},\,  k\in \Z,\, i=1,\ldots,N\},
\ee
and
\be
\dicw_{j}=
\{2^{j/2}\psi(2^jx_i-\frac{k}{2})|_{[c,d]},\,  k\in \Z,\, i=1,\ldots,N\},
\ee
where $\psi(2^jx_i-\frac{k}{2})|_{[c,d]}$ indicates the restriction 
of the function  $\psi(2^jx_i-\frac{k}{2})$ to the interval $[c,d]$.
{{Different families of wavelet basis and dictionaries for 
approximation of heartbeats have been compared in 
\cite{RNC19}, 
where the Cohen-Daubechies-Feauveau family was singled out as 
the most effective one. We have confirmed the same outcome
for simultaneous approximations and adopted the 
 Cohen-Daubechies-Feauveau $\CDF$ 
 dictionary of redundancy approximately two 
 introduced in \cite{RNC19}.}}

The prototype functions $\phi(x)$ and $\psi(x)$  
are plotted in the left and right 
graphs of Fig.\ref{wavelets} respectively. 
The MATLAB codes for producing numerically both functions 
and building 
the dictionary \eqref{dic} are described in \cite{CRN20}. The codes 
have been made 
available in \cite{webpage} together with of the complete MATLAB 
software for reproducing the numerical examples in this work. 

\begin{figure}[ht!]
\begin{center}
\includegraphics[width=6cm]{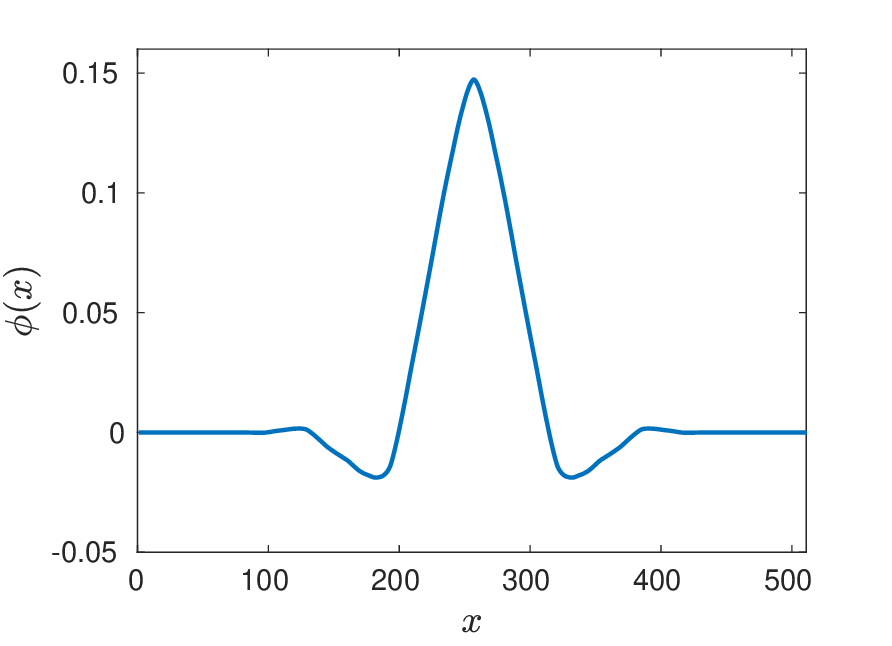}
\includegraphics[width=6cm]{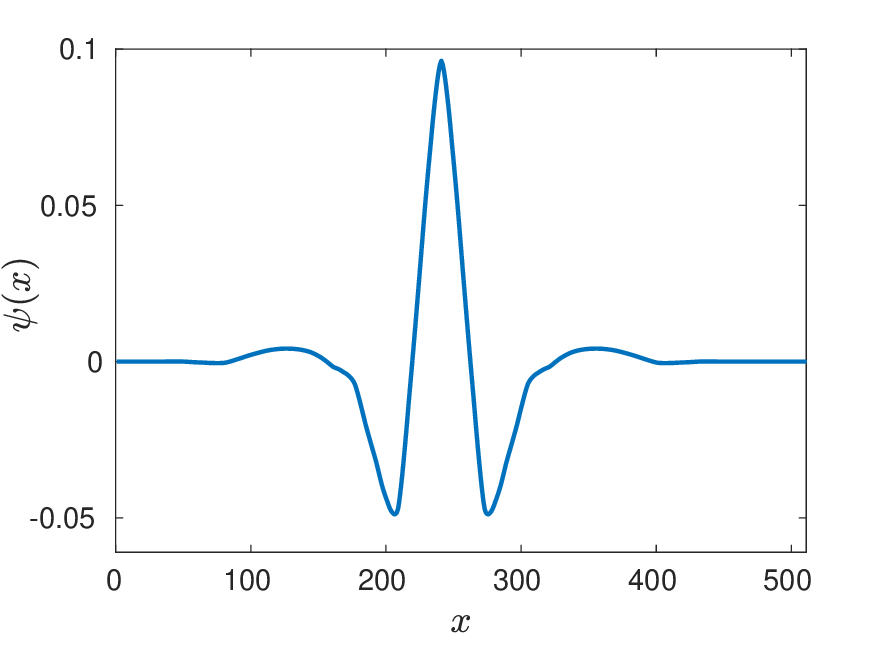}
\end{center}
	\caption{Cohen-Daubechies-Feauveau 
	scaling and wavelet functions \cite{CDF92}.}
\label{wavelets}
\end{figure}

{{
The segmented and aligned heartbeats are simultaneously
approximated using the SOOMP approach by assigning the same
weight to each heartbeat, i.e. 
$p(q)=\frac{1}{Q},\,q=1,\ldots,Q$. In this case the algorithm 
stops at iteration $k$ if  
$$\sum_{q=1}^Q\frac{1}{Q} \|\vfq -\vfq^k\| < \rho,$$
with
\be
\rho=\frac{\PRDN_0}{100}\sum_{q=1}^Q\frac{1}{Q} \|\vfq -\overline{\vfq}\|,
\ee
$\overline{\vfq}$ indicating the mean of $\vfq$  and 
$\PRDN_0$ a fixed value of the metric of quality 
$\PRDN$, which is defined by  
\be
\label{prdn}
\PRDN= \frac{\|\vf - \vf^a\|}{\|\vf - \overline{\vf}\|} \times 100\%, 
\ee
where $\vf$ is the whole ECG record, $\vf^a$ is the reconstructed 
record from the approximated heartbeats  
and ${\overline{\vf}}$ is the mean of $\vf$. 

Given a required value of $\PRDN$, at the approximation step of 
the processing the 
parameter $\PRDN_0$
is fixed as $0.8 \cdot\PRDN$, in order to achieve the 
target value $\PRDN$ at the quantization 
step described in subsection \ref{encoding}.

Even if the approximation is realized to 
 achieve the required $\PRDN_0$ by the whole record, 
it is  
interesting to calculate the quality metric for each heartbeat 
in the array. 
To this end we defined  
\be
\prdn(q)= \frac{\|\vfq - \vf^a\{q\}\|}{\|\vfq - \overline{\vfq}\|} \times 100\%,\quad q=1,\ldots,Q,
\ee
where $\vf^a\{q\}$ is the approximation of the beat $\vfq$
and $\overline{\vfq}$ its mean value.
The values of $\prdn(q)$ for the simultaneous 
	approximation of records  111 and 100 
are shown in the left and right graphs of Fig.\ref{pdrnq} 
respectively.
\begin{figure}[ht!]
\begin{center}
\includegraphics[width=6cm]{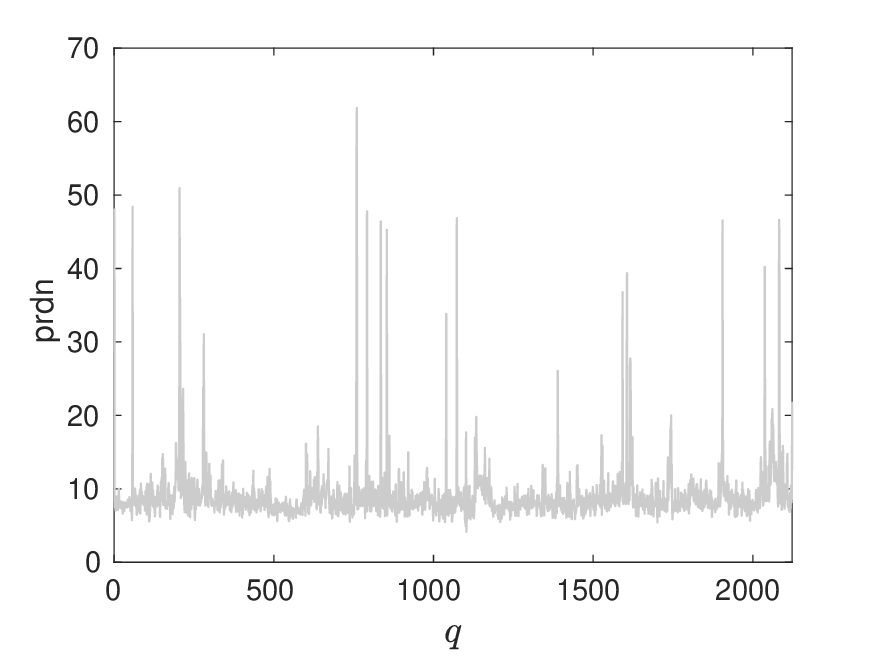}
\includegraphics[width=6cm]{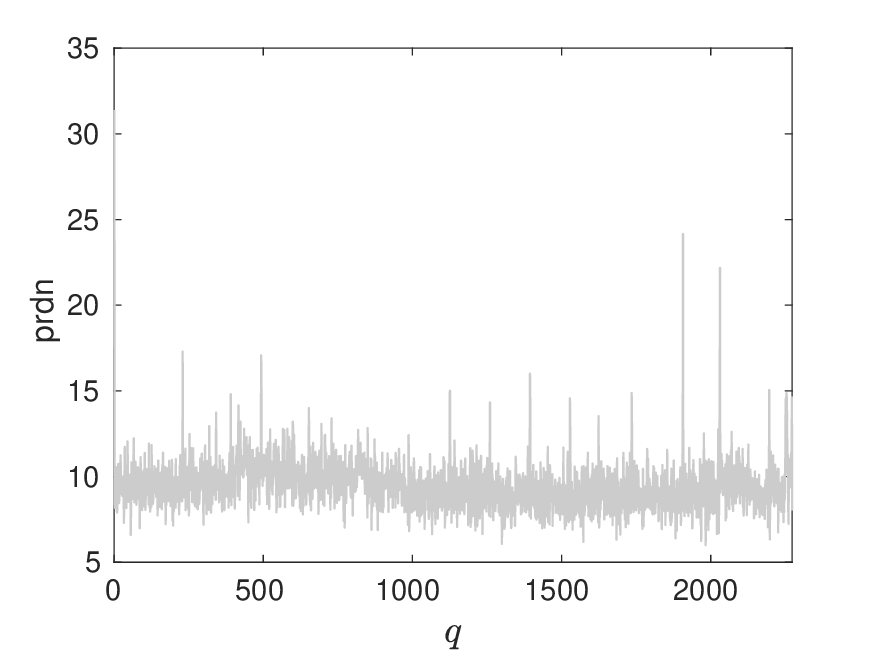}
\end{center}
	\caption{Values of prdn from the simultaneous 
	approximation of the aligned beats in 
	records 111 (left graph) and 100 (right graph).}
\label{pdrnq}
\end{figure}
The total $\PRDN$ produced by
the reconstruction of the whole records is, 
in both cases, $\PRDN=9.1$. The mean value 
$\prdn$ for record 111 is 9.0 with std=4 while for record 100 
 the mean $\prdn$ is 9.4 with std=1.4.
 However, as noticeable 
in the figures, for some $q$s the 
 $\prdn$ values are much higher than for others. 
This is a consequence of 
the irregularities of the beats, which can 
be perceived in the left image of Fig.~\ref{2Dbeats}. 

\begin{figure}[ht!]
\begin{center}
\includegraphics[width=6cm]{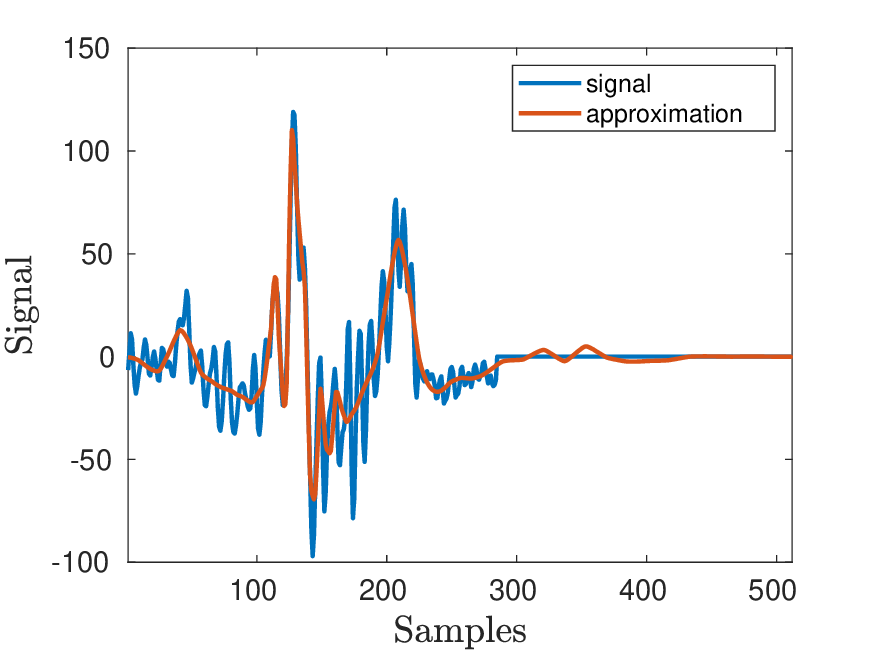}
\includegraphics[width=6cm]{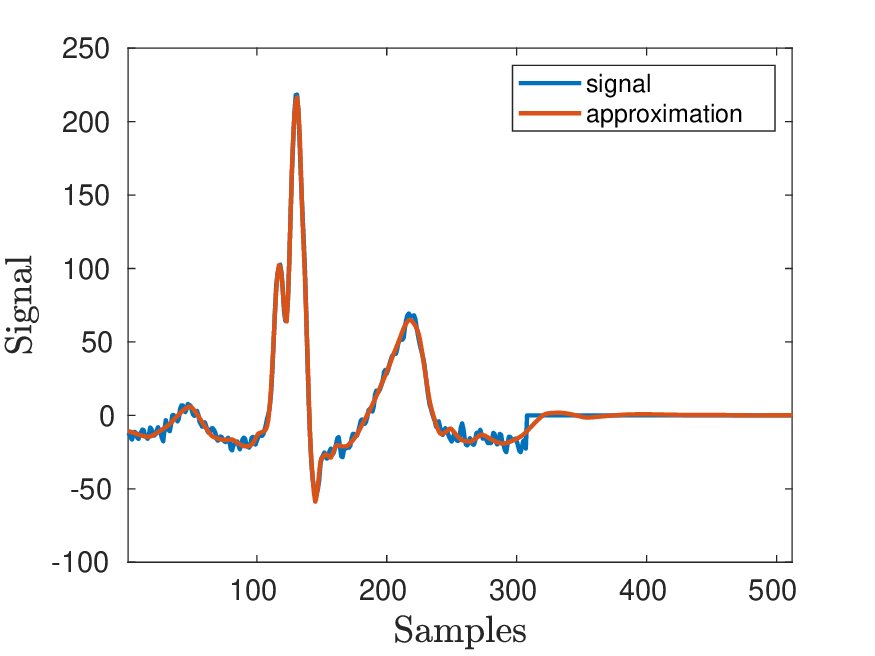}\\
\includegraphics[width=6cm]{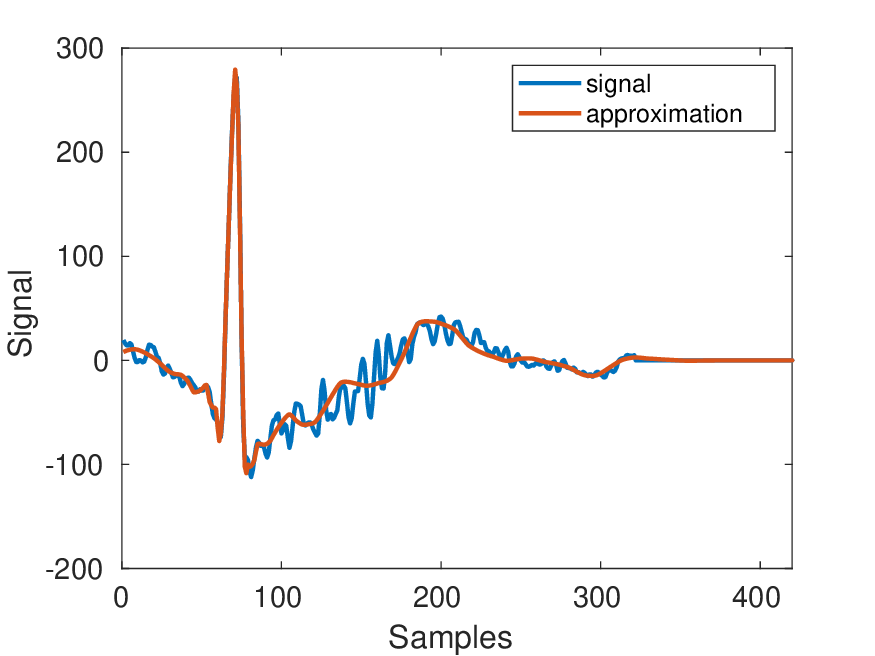}
\includegraphics[width=6cm]{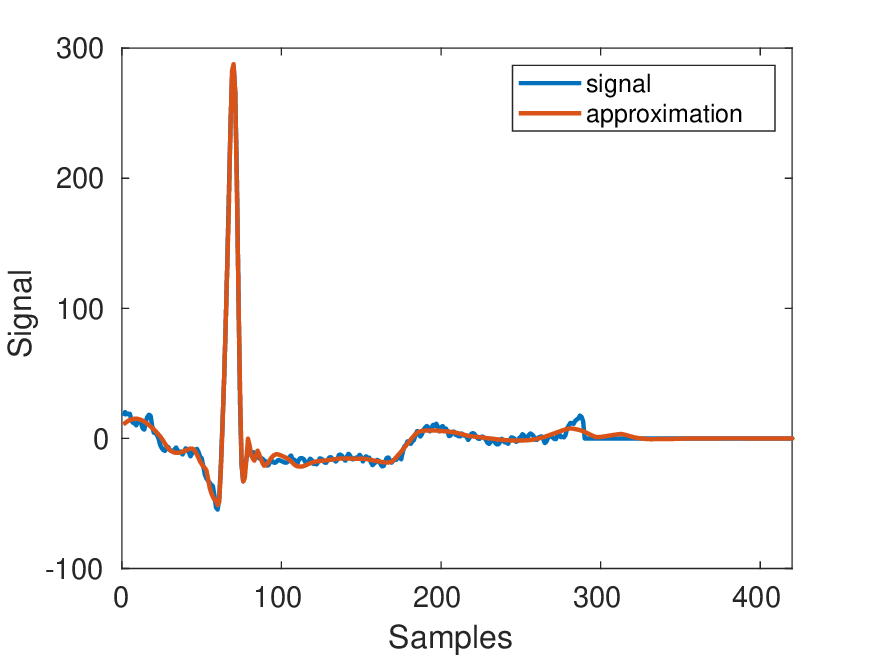}\\
\end{center}
        \caption{The top left graph shows  one of
        the heartbeats in records 111, and its approximation, which
        yields an outlier value of $\prdn$. The
        right graph corresponds to a beat  yielding $\prdn$ close to
        the mean value $\prdn=9$. The bottom graphs have the same
        description but the heartbeats are from record 100.}
\label{beat_app}
\end{figure}

The top left graph of Fig.~\ref{beat_app} depicts a 
	heartbeat in records 111 and 
its approximation. This heartbeat yields $\prdn=46.5$.  
As shown by the red line the figure,  such 
high $\prdn$ value is produced by a 
smooth version of the noisy signal. 
On the contrary, for regular heartbeats  the
 $\prdn$ values are close to the 
$\PRDN$ of the whole record. The right graph of the figure 
shows the approximation of one of those beats. 
The left bottom graph is the approximation of one of the few
beats in record 100 which yields the outlier value  $\prdn=20$.
The right bottom graph shows the approximation of one of the other
beats. 	

By the simultaneous approximation of the aligned heartbeats
these are transformed into a reduced set of numbers which 
allow to reconstruct the approximated heartbeats. 
This set consists of 
a) the $k$ indices $\ell_n,\,n=1,\ldots,k$  corresponding to the common atoms in the decomposition of the heartbeats (c.f. \eqref{atoq}) 
b) the different coefficients $c \cq(n),\,n=1,\ldots,k, \, q=1,\ldots,Q$  in the decomposition of each heartbeat (c.f. \eqref{atoq}). 
These coefficients can be 
	placed in a two dimensional array $\vC \in \R^{Q \times k}$
        as illustrated in Fig.~\ref{2Dcoe}.
	The top image on the left in this figure 
	shows the magnitude of the 
	array $\vC \in \R^{2133 \times 33}$ arising from 
	the approximation up to $\PRDN=9$ of record.
	111. The bottom left image corresponds to the approximation 
	up to $\PRDN=9$ of record 100.

\begin{figure}[ht!]
\begin{center}
\includegraphics[width=6cm]{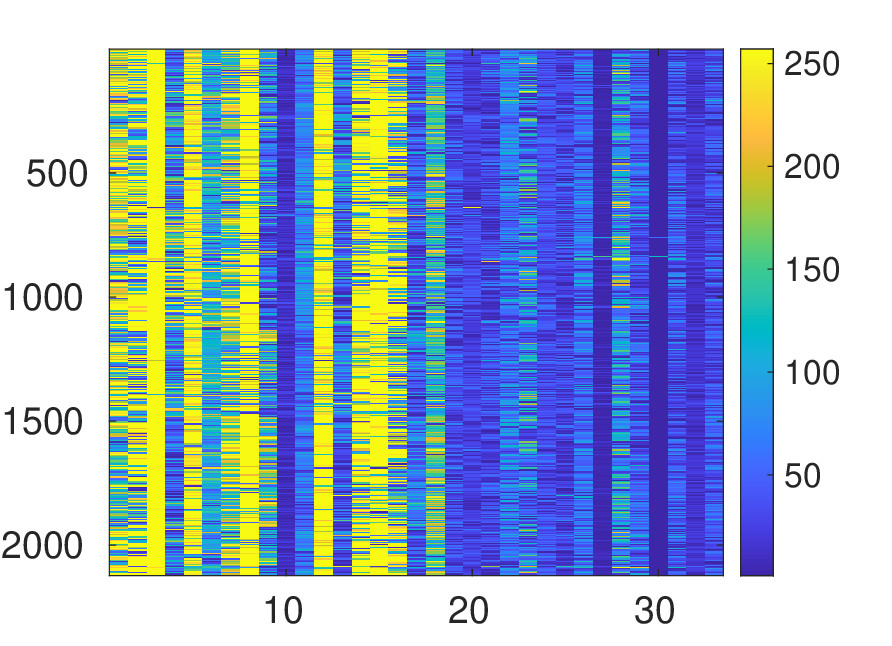}
\includegraphics[width=6cm, height=4.31cm]{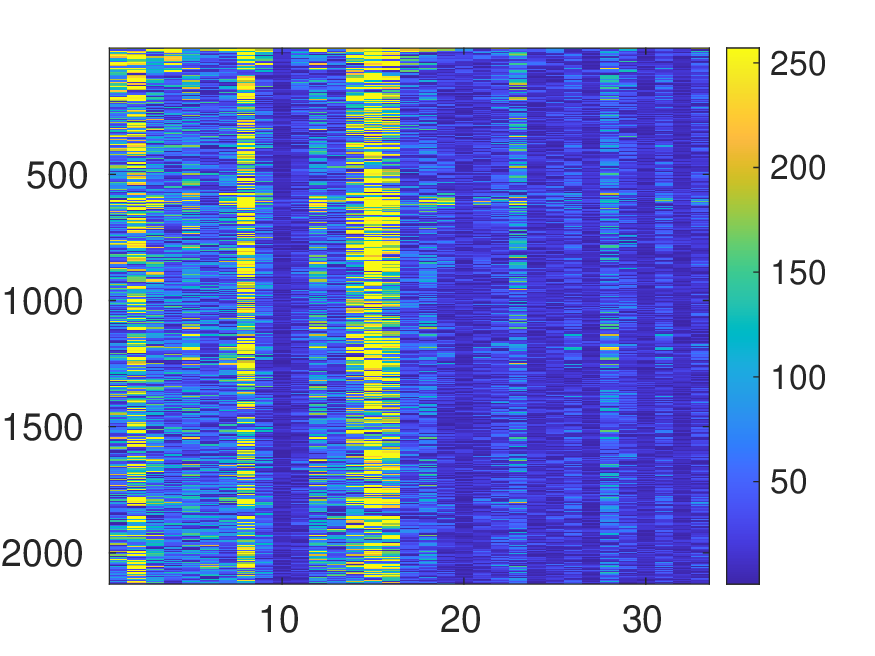}\\
\includegraphics[width=6cm]{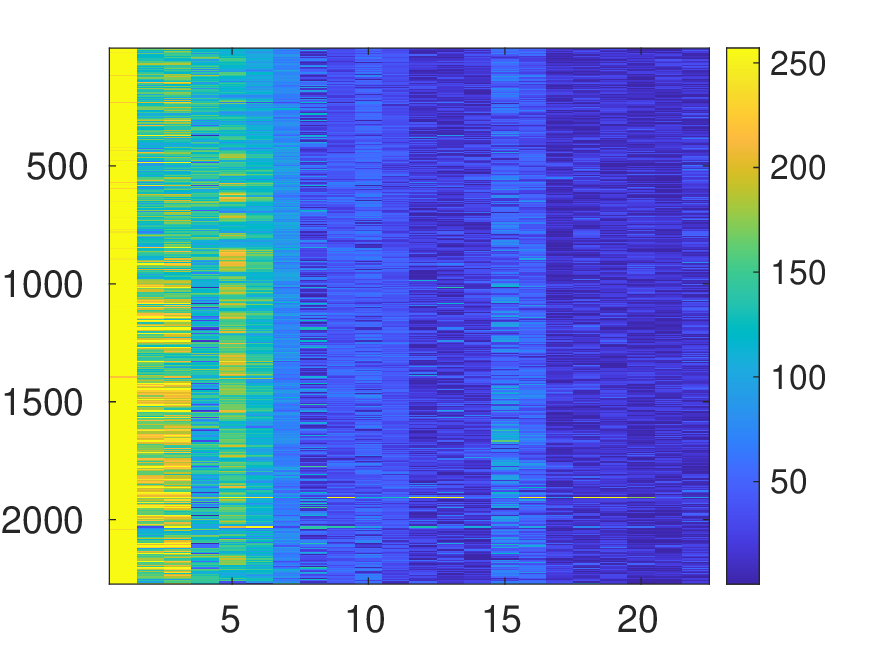}
\includegraphics[width=6cm]{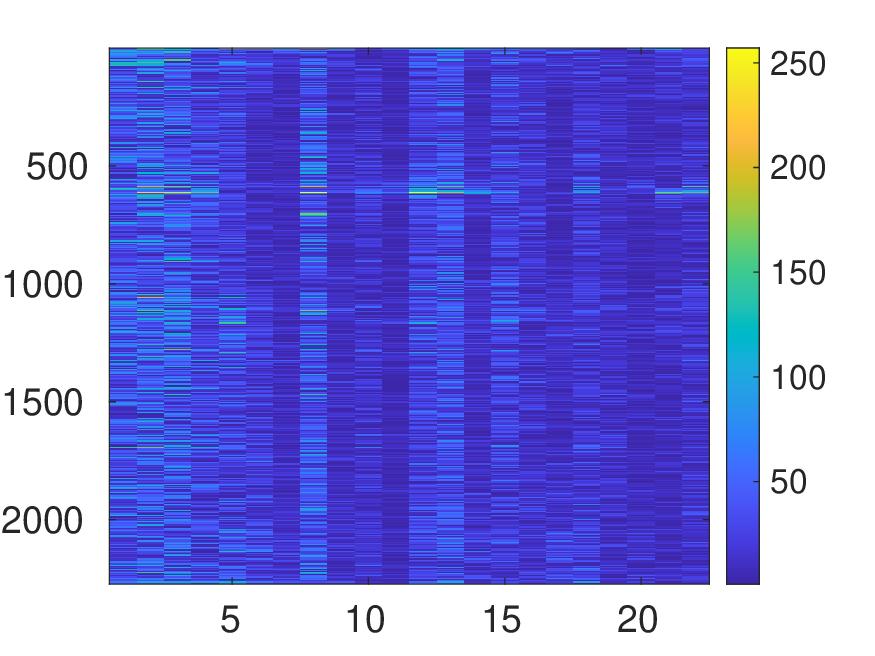}
\end{center}
        \caption{The top image on the left depicts the 
	magnitude of entries in the array 
	$\vC \in \R^{2133 \times 33}$ containing the 
	coefficients in the approximation of record 
	111. The bottom left image depicts the magnitude of entries in the arrays $\vC \in \R^{2273 \times 25}$ corresponding to 
	the approximation of records 100. 
	The images on the right 
	are the magnitude of the entries in 
	the arrays $\vCt \in \R^{2133\times 33}$ 
	(top) and  $\vCt \in \R^{2273\times 25}$ (bottom) 
	arising by applying the discrete cosine transform 
	on the columns of the arrays represented by 
	the images on the left.}
\label{2Dcoe}
\end{figure}

It is clear from the location of the brightest pixels 
in the left images of Fig.~\ref{2Dcoe} that the coefficients
of largest
magnitude are concentrated in vertical lines.
This suggests that, 
to favor compression for storing these values,  
it is convenient to apply an orthogonal
transformation to map the coefficients in the vertical
direction to smaller values which eventually might be quantized
to zero. Consequently, 
by applying the discrete cosine transform on each
  column of $\vC$, we create the transformed array 
  $\vCt \in \R^{Q\times k}$ with the following entries
\be
\label{At}
\vCt(:,n)= \DCT \vC(:,n),\quad n=1,\ldots,k,
\ee
where $\DCT \vC (:,n)$ indicates the one dimensional
 discrete cosine transform operating on the $n$-th column of
array $\vC$.
The transformed points corresponding to the 
left images in Fig.~\ref{2Dcoe}  
are represented in the right images of this figure. 
The introduction of this step to decorrelate the 
vertical entries in the array $\vC$ is 
 key to boost the performance of the 
adopted encoding strategy described below. 
The notorious change of intensity in the images on 
the right of Fig.~\ref{2Dcoe} indicates that 
after quantization some of the  entries of the transformed
arrays will be mapped to zero. Within the 
encoding strategy described in the next section, 
this effect enhances compression. 
}}

\subsection{Encoding}
\label{encoding}
 At the encoding step the  $Q \times k$
 array $\vCt$ is expressed as a 
 vector $\vat=(\at(1),\ldots,\at(K))$ 
 of $K=Q\cdot k$ components, 
 adopting the column-major order. The encoding of 
 this vector follows the procedure outline in \cite{LRN19}.
 The components of $\vat$
 are converted to integer numbers
 by a mid-tread uniform quantizer as follows:
\be
\label{uniq}
\at^\Delta(i)= \left \lfloor \frac{\at(i)}{\Delta} +\frac{1}{2} \right \rfloor,\quad  i=1,\ldots,K,
\ee
where $\lfloor x \rfloor$ indicates the largest
integer
smaller or equal to $x$ and $\Delta$ is the quantization
parameter. For comparison with results in other publications   
in the numerical examples 
the quantization 
parameter $\Delta$ is set 
to produce the required quality of the reconstructed signal. 

The absolute value of
the elements  \eqref{uniq} are placed in
a smaller vector, say $\vbp= (b'(1),\ldots,b'(K'))$, after the
elimination of zeros. The signs are encoded
separately in a vector $\vs=(s(1),\ldots,s(K'))$
 using a binary alphabet: 1 for $+$ and 0 for $-$.

Assuming that the nonzero values in \eqref{uniq} occur at
the positions $\jll_i,\ldots,\jll_{K'}$, these
indices are re-ordered in ascending
 order $\jll_{i} \rightarrow \til{\jll}_i,\,i=1,\ldots,K'$,
 i.e. 
$\til{\jll}_i < \til{\jll}_{i+1},\,i=1,\ldots,K'$.
This induces new order in the coefficients,
$\vbp \rightarrow \vbpt$ and in the corresponding 
signs $\vs \rightarrow
\vst$.
Defining $\delta(i)=\jllt_i-\jllt_{i-1},\,i=2,\ldots,K'$
the array
$\vdel=(\jllt_1, \delta(2), \ldots, \delta(K'))$ stores
 the indices
$\jllt_1, \ldots, \jllt_{K'}$ with unique recovery.

Finally the vectors $\vbpt, \vst, \vdel$, as well as
the length of the
heartbeats $\vhb$, are compressed  using adaptive Huffman
coding implemented by the
 off-the-shelf MATLAB function Huff06 \cite{Karl}.
 The additional numbers which have to be
passed to the decoder are:\\
(i) The indices
$\ell_i, \,i=1,\ldots,k$ of the
selected dictionary's atoms forming the common basis.\\
(ii) The quantization parameter
$\Delta$.\\
(iii) The mean value of the 1D ECG record (if not previously subtracted).\\
(iv) The number of rows and columns of $\vC$, i.e. $Q$ and $k$.\\

\subsection{1D ECG signal recovery}
At the decoding stage, after reverting
 Huffman coding, the locations $\jllt_1, \ldots, \jllt_K$
of the nonzero entries in the transformed array 
after quantization 
 are readily obtained. This allows the
recovery of the array $\vAtr$ as follows.
\begin{itemize}
	\item[(i)] Set $\atr(i)=0,\, i=1,\ldots, K$  and
$\atr(\jllt_i)= (2\st(i)-1)\bpt(i)\Delta,\, i=1,\ldots,K$.
\item[(ii)] Reshape the 
 vector $\vatr$ to produce a 2D array $\vAtr$ of size $Q \times k$.
The array $\vCr$ is recovered from the $\vAtr$ one by inverting
		the $\DCT$ transformation (c.f.  \eqref{At}). 
	\item[(iii)] Each row of the recovered array $\vCr$ gives the 
	coefficients in the decomposition  \eqref{atoq} of 
		the approximated heartbeats, i.e.
		$\vfr\{q\}= \sum_{i=1}^k \vCr(q,i) \vd_{\ell_i}, \quad
		q=1,\ldots,Q.$
	\item[(iv)] Finally the reconstructed beats $\vfr\{q\}$ are assembled 
	in a 1D record using the distance between heartbeats 
 that was stored in the vector $\vhb$.
\end{itemize}

The achieved compression ratio $\CR$, which is defined as
\be
\text{CR}=\frac{{\text{Size of the uncompressed file}}}{{
\text{Size of the compressed file}}},
\ee
depends on the required quality of the
recovered signal. In the numerical examples the quality of the 
recovered records is assessed by the $\PRDN$ as  defined  in 
\eqref{prdn}. It is pertinent to stress the importance of 
 adopting this normalized  metric 
 for comparison of reconstruction quality. 
The subtraction of $\overline{\vf}$ avoids
 dependence on the signal baseline.  
\section{Numerical Tests}
For the numerical test we use the MIT-BIH Arrhythmia 
database \cite{MITDB}. 
Each of the records is of
30 min length, consisting of
 $N=650000$ 11-bit samples at a frequency of 360 Hz.

 For comparison purposes 
 we compress the subset of records 
 reported in \cite{AGL15}, \cite{LLW18}, and \cite{ZWL21} and 
 reproduce the values of $\PRDN$ in
 those publications.  This is achieved as follows: 
 The SOOMP method is applied to approximate the set of 
 heartbeats in each record up to 80\% the target $\PRDN$. 
 The quantization parameter $\Delta$ is automatically 
 fixed, by a bisection algorithm, 
 in order to reproduce the target $\PRDN$ for the whole record 
 within two decimal places.

 The first, second and third columns of
 Table~\ref{TABLE1} reproduce the results published in 
 \cite{AGL15}. 
 The comparison is relevant because the 
 approach \cite{AGL15} is also based on approximation of heartbeats using a 
 dictionary.  
 The techniques are very different though.  Whilst our
 dictionary does not have to be stored because it is 
  numerically generated, 
 the dictionary in \cite{AGL15} 
 is part of the ECG record to be compressed. 
 Moreover, the method for finding the 
 sparse representation is different and so is the  
 procedure to store the parameters that 
 should be passed to the decoder. 

\begin{table}[h!]
\caption{Comparison with the results in \cite{AGL15}. 
The first collumn 
lists the records considered in \cite{AGL15}. 
The second column displays  
the values of $\PRDN$ and the third collum 
their $\CRs$. Our $\CRs$ for the same $\PRDN$ are shown in the forth column. 
The fifth column shows the 
$\CRs$ obtained with the fast approach \cite{LRN19}.} 
\label{TABLE1}
\begin{center}
\begin{tabular}{||c r r r r||}
\hline \hline
Record &$\PRDN$&$\CR$\cite{AGL15}& $\CR$ prop. &$\CR$
 \cite{LRN19}
\\ \hline \hline

100&18.03 & 78.20&  {\bf{143.99}}&  36.51  \\
	100&17.22  & 75.12 & {\bf{139.47}} &  35.25   \\
 &  & & & \\
	101  &   14.66 &  80.24&  {\bf{102.58}}&  31.26   \\ 
	101  &   12.91 &  76.46&  {\bf{82.31}}&   30.31 \\
&  & & &\\
	102  &   18.54 &  {\bf{58.54}}&  {\bf{58.49}}&    33.89\\
	102  &   18.16 &  48.47&  {\bf{58.13}}&    33.48\\
&  & & &\\
	103  &   12.57 &  46.32&  {\bf{90.91}}&   30.84 \\  
	103  &   11.57 &  44.33&  {\bf{86.27}}&    29.61 \\
&  & & &\\
	109  &   13.70 &  24.86&  {\bf{145.80}}&    51.23 \\
	109  &    9.97 &  23.53&   {\bf{97.73}}&   36.91\\
&  & & &\\
	111  &   26.20 &  31.05&   {\bf{121.09}}&  38.29  \\ 
	111  &   19.51 &  29.44&    {\bf{60.49}}&  32.20 \\ 
	&  & & &\\
	112  &   16.58 &  34.06&   {\bf{91.48}}&    35.05 \\  
	112  &   15.99 &  35.49&   {\bf{85.59}}&    34.32\\ 
	&  & & &\\
	113  &   14.08 &  37.42&   {\bf{90.76}}&   32.49  \\
	113  &    9.82 &  32.55&   {\bf{55.30}}&   27.68 \\
	&  & & &\\
	115  &   9.76  &  38.26&   {\bf{62.31}}&   24.52  \\ 
	115  &   9.18  &  36.57&   {\bf{57.32}}&   23.74 \\
	&  & & &\\
	117  &   14.42 &  38.94&   {\bf{120.89}}&  36.94  \\  
	117  &   13.38 &  37.13&   {\bf{105.97}}&   35.74  \\  
	&  & & &\\
	119  &   32.19 &  16.26&   {\bf{153.33}}&  90.40  \\ 
	119  &   16.36 &  15.24&   {\bf{78.81}}&   48.08 \\ 
	&  & & &\\
	121  &   17.36 &  26.67&   {\bf{111.74}}&   46.45   \\ 
	121  &   15.63 &  25.29&    {\bf{100.72}}&   41.11  \\
	&  & & &\\
	Average& 17.33&  41.9\;  &  {\bf{107.78}}&  40.65\\
	Average& 14.14&  39.97 &  {\bf{84.00}} &  34.04\\ \hline\hline
\end{tabular}
\end{center}
\end{table}

 Our compression results are shown in the forth column of 
 Table~\ref{TABLE1}. These results demonstrate a significant gain in 
 $\CR$ for the same recovery quality. For further comparison 
 we apply the fast compression algorithm \cite{LRN19}, which does not require peak segmentation or Huffman coding. This method  
 has been already shown 
 to improve the average $\CR$ for the 48 records 
 in the MIT-BIH Arrhythmia dataset with respect to the results in 
 \cite{LKL11}, \cite{MZD15}, 
 and \cite{TZW18},
 for a broad rage of average qualities. For comparison with 
 \cite{AGL15} in Table~\ref{TABLE1} the compression is realized 
 to reproduce the $\PRDN$ listed in the second column for each 
 record.
 %
 
\begin{table}[h!]
	\caption{Same description as in Table \ref{TABLE1} but 
	the comparison is with the results of Table I in \cite{LLW18}.}
\label{TABLE2}
\begin{center}
\begin{tabular}{||c r r r r||}
\hline \hline
Record &$\PRDN$&$\CR$\cite{LLW18}& $\CR$ prop. &$\CR$
 \cite{LRN19}
\\ \hline \hline

	100  &   11.46  &    39.81  &  {\bf{64.47}} &  23.03 \\

	101  &   14.13  &    42.04  &   {\bf{95.53}} &   30.91\\

	102  &   19.94  &    41.09  &   {\bf{63.69}}  & 35.09\\

	103  &   6.72   & {\bf{41.24}} &  39.17  & 21.05\\

	107  &   13.27  &    41.84  &   {\bf{71.40}}  & 38.15\\

	109  &   7.31   &    38.25  &   {\bf{59.34}}  & 28.46\\

	111  &   13.94  &  {\bf{41.73}} & {\bf{41.59}}  & 25.14\\

	115  &   8.04   &     42.71 &    {\bf{47.79}}   & 22.04\\

	117  &   10.00  &     46.75 &    {\bf{51.56}}   & 26.17\\

	118  &   15.33  &     39.60 &    {\bf{66.44}}   & 28.26\\

	119  &   9.67   &     41.97 &    {\bf{43.98}}   & 30.71\\

	213  &   13.63  &     32.58 &    {\bf{62.24}}   & 25.72\\

	222  &   22.44  &     {\bf{40.69}} &  31.95   &24.77\\

	232  &   20.76  &     42.28 &    {\bf{59.66}}   &35.52\\

	Average &  13.33 & 40.90  &    {\bf{57.09}}   & 28.22\\ \hline \hline
\end{tabular}
\end{center}
\end{table}

The first, second and third columns of
 Table \ref{TABLE2} reproduce the results published in \cite{LLW18}, 
 which are achieved with an approach based 
 on the Singular Value Decomposition (SVD).

 Our compression ratios ($\CRs$) are shown in the 
  the forth column of Table \ref{TABLE2}. The fifth column shows the 
  $\CRs$ produced by the fast compression algorithm \cite{LRN19}.

\begin{table}[h!]
	\caption{Same description as in Table \ref{TABLE1} but 
	the comparison is with the results of Table I in \cite{ZWL21}.}
\label{TABLE3}
\begin{center}
\begin{tabular}{||c r r r r||}
\hline \hline
Record &$\PRDN$&$\CR$\cite{ZWL21}& $\CR$ prop. &$\CR$
 \cite{LRN19}
\\ \hline \hline
	100 &  11.55 &  50.70 &  {\bf{65.47}}  &23.14\\
	101 &  11.29 &  58.54 &  {\bf{64.13}}  &27.86\\
	103 &   9.16 &  54.16 &  {\bf{61.37}}  &25.70\\
	107 &  14.53 &  53.12 &  {\bf{78.31}}  &40.34\\
	109 &  11.83 &  46.43 &  {\bf{120.08}} &44.67\\
	111 &  16.40 &  {\bf{53.39}} & 50.11 &28.34\\
	115 &   8.94 &  56.77 &  {\bf{58.76}} &23.49\\
	117 &  12.43 &  66.15 &  {\bf{90.39}}  &34.62\\
	119 &  10.28 &  {\bf{56.03}} & 46.99  &32.17\\
	214 &  17.03 &  50.84 &  {\bf{87.66}}  &54.04\\
	223 &  17.49 &  45.38 &  {\bf{86.06}}  &40.80\\
	Avegave & 12.81& 53.77 & {\bf{73.57}}  &34.11\\
\hline \hline

\end{tabular}
\end{center}
\end{table}

The first, second and third columns of
 Table \ref{TABLE3} reproduce the results published in
 \cite{ZWL21}, 
 which are also obtained with 
 a Singular Value Decomposition based approach. 
Our $\CRs$ are shown in the forth column of this table.
The fifth column shows the
  $\CRs$ produced by the fast compression algorithm \cite{LRN19}.

Note: 
The MATLAB software for reproducing the tables is available on\\
{\tt{http://www.nonlinear-approx.info/examples/node017.html}}

\section{Conclusions}

The Optimized Orthogonal Matching Pursuit approach has been 
extended with the purpose of selecting a common basis for 
the simultaneous approximation of a set of 
similar signals. The extended approach, termed
Simultaneous Optimized Orthogonal Matching Pursuit, 
minimizes at each iteration the mean value  square 
error norm of the joint approximation. 
{{The algorithm's implementation was demonstrated by 
 approximating 
stereophonic music using 
 a highly coherent trigonometric dictionary.}} The applicability of the method to ECG compression was illustrated 
on records taken from the MIT-BIH Arrhythmia database. 
The particular records 
were selected for comparison purposes as in 
\cite{AGL15}, \cite{LLW18}, and \cite{ZWL21}.
The simultaneous approximations of aligned 
 heartbeats was used for compressing a whole record.
The adopted compression strategy was shown to improve upon
compression results achieved by other methods for the same
reconstruction quality. The comparison was made possible
by means of an iterative quantization procedure which delivers
the required quality.

{{
While the proposed approach 
involves detection and alignment of R-peaks,
it is the approximation step which introduces the highest
computational cost. In order to address this matter as a line
of future work, it would be interesting to
investigate the possibility of selecting a suitable 
subspace $V_k$ from 
a whole data set (instead of a subspace for each single
record as is done here).
Certainly approximating  new
records using  previously selected atoms  would
significantly speed the compression procedure. It is
still to be  discerned if the SOOMP approach
could pick out a common low dimension  
subspace to approximate, up to a given quality, 
any new ECG record.
We feel confident that the results presented in this work 
 will motivate further research in the topic.
}}

\end{document}